\documentclass[aps,prd,preprint,preprintnumbers,nofootinbib,superscriptaddress]{revtex4}
\usepackage{ae}
\usepackage{bm} 
\usepackage{color}
\usepackage{amsmath}
\usepackage{amssymb}
\usepackage{amsfonts}
\usepackage{graphicx}
\usepackage{setspace}
\usepackage{stackrel}
\usepackage{ulem}
\usepackage{braket}
\usepackage{xfrac}
\usepackage{siunitx}
\usepackage{upgreek}
\usepackage{multirow}

\graphicspath{{./plots/}}

\newcommand{\Eqref}[1]{Eq.~\eqref{#1}}

\allowdisplaybreaks

\begin{document}

\setlength{\unitlength}{1mm}
\title{A dark-field setup for the measurement of light-by-light scattering with high-intensity lasers}
\author{Fabian Sch\"utze}\email{fabian.schuetze@uni-jena.de}
\affiliation{Helmholtz-Institut Jena, Fr\"obelstieg 3, 07743 Jena, Germany}
\affiliation{GSI Helmholtzzentrum f\"ur Schwerionenforschung, Planckstra\ss e 1, 64291 Darmstadt, Germany}
\affiliation{Theoretisch-Physikalisches Institut, Abbe Center of Photonics, \\ Friedrich-Schiller-Universit\"at Jena, Max-Wien-Platz 1, 07743 Jena, Germany}
\author{Leonard Doyle}
\affiliation{Ludwig-Maximilians-Universit\"at M\"unchen, Am Coulombwall 1, 85748 Garching, Germany}
\author{J\"org~Schreiber}
\affiliation{Ludwig-Maximilians-Universit\"at M\"unchen, Am Coulombwall 1, 85748 Garching, Germany}
\author{Matt Zepf}
\affiliation{Helmholtz-Institut Jena, Fr\"obelstieg 3, 07743 Jena, Germany}
\affiliation{GSI Helmholtzzentrum f\"ur Schwerionenforschung, Planckstra\ss e 1, 64291 Darmstadt, Germany}
\affiliation{Theoretisch-Physikalisches Institut, Abbe Center of Photonics, \\ Friedrich-Schiller-Universit\"at Jena, Max-Wien-Platz 1, 07743 Jena, Germany}
\author{Felix Karbstein}\email{felix.karbstein@uni-jena.de}
\affiliation{Helmholtz-Institut Jena, Fr\"obelstieg 3, 07743 Jena, Germany}
\affiliation{GSI Helmholtzzentrum f\"ur Schwerionenforschung, Planckstra\ss e 1, 64291 Darmstadt, Germany}
\affiliation{Theoretisch-Physikalisches Institut, Abbe Center of Photonics, \\ Friedrich-Schiller-Universit\"at Jena, Max-Wien-Platz 1, 07743 Jena, Germany}

\date{\today}

\begin{abstract}
We put forward a concrete experimental setup allowing to measure light-by-light scattering in the collision of two optical high-intensity laser beams at state-of-the-art high-field facilities operating petawatt class laser systems.
Our setup uses the same focusing optics for both laser beams to be collided and employs a dark-field approach for the detection of the single-photon-level nonlinear quantum vacuum response in the presence of a large background.
Based on an advanced modeling of the colliding laser fields, we in particular provide reliable estimates for the prospective numbers of signal photons scattered into the dark-field for various laser polarizations.
\end{abstract}

\maketitle

\section{Introduction}\label{sec:intro}

Quantum fluctuations mediate effective couplings of macroscopic electromagnetic fields in vacuum. These supplement Maxwell's linear theory of classical electrodynamics with effective nonlinear interactions \cite{Heisenberg:1935qt,Weisskopf:1936hya,Schwinger:1951nm}.
All macroscopic electric $\vec{E}$ and magnetic $\vec{B}$ fields currently available in the laboratory meet the criterion $\{ |\vec{E}|, c|\vec{B}| \} \ll E_{\rm cr}$, with the reference field strength $E_{\rm S} = m^2c^3/(e\hbar)\simeq 1.3 \times 10^{18}\,{\rm V}/{\rm m}$ set by the electron mass $m$ and elementary charge $e$, respectively.
If these fields vary on spatial and temporal scales $\lambda,\tau$ much larger then the Compton wavelength of the electron $\lambdabar_{\rm C}=\hbar/(mc) \simeq 3.9 \times 10^{-13}\,{\rm m}$, i.e., fulfill the low-energy condition $\{\lambda,c\tau\}\gg\lambdabar_{\rm C}$, the leading quantum vacuum non-linearity can be parameterized by the interaction Lagrangian ($\hbar=c=1$) \cite{Euler:1935qgl}
\begin{equation}
    {\cal L}_{\rm int} \simeq \frac{m^4}{1440\pi^2}\Biggl[a \biggl(\frac{\vec{B}^2-\vec{E}^2}{E_{\rm S}^2}\biggr)^2+b\biggl(\frac{2\vec{B}\cdot\vec{E}}{E_{\rm S}^2}\biggr)^2\Biggr]
    =\frac{1}{360\pi^2}\Bigl(\frac{e}{m}\Bigr)^4\bigl(a{\cal F}^2+b{\cal G}^2\bigr)\,,
    \label{eq:HElagrangian}
\end{equation}
where the numerical constants $a$ and $b$ are fully determined by the underlying quantum field theory; ${\cal F}=F_{\mu\nu}F^{\mu\nu}/4$ and ${\cal G}=F_{\mu\nu}{}^\star\!F^{\mu\nu}/4$ with (dual) field strength tensor $F^{\mu\nu}$ (${}^\star\!F^{\mu\nu}$).

Within the standard model of particle physics their values are accurately determined by quantum electrodynamics (QED), predicting these to be given by \cite{Euler:1935zz,Ritus:1975pcc,Gies:2016yaa}
\begin{equation}
a=4\Bigl(1+\frac{40}{9}\frac{\alpha}{\pi}+{\cal O}(\alpha^2)\Bigr)\ \ \ \text{and}\ \ \
b=7\Bigl(1+\frac{1315}{252}\frac{\alpha}{\pi}+{\cal O}(\alpha^2)\Bigr)\,, \label{eq:a,b}
\end{equation}
where $\alpha=e^2/(4\pi)\simeq1/137$ is the fine structure constant.
The above {\it weak} and {\it slowly varying} field conditions are in particular well met by the field configurations generated by state-of-the-art high-intensity lasers that produce the strongest macroscopic electromagnetic fields currently available in the laboratory.
Higher-order QED corrections to \Eqref{eq:HElagrangian} are parametrically suppressed by additional powers of $1/E_{\rm S}\sim\lambdabar_C^2\sim1/m^2$.

The effective couplings of macroscopic electromagnetic fields in \Eqref{eq:HElagrangian} generically gives rise to a signal component that may differ in characteristic properties such as propagation direction and polarization from the originally applied fields. Because these signals
are very small, they could not yet be verified in a controlled laboratory experiment \cite{Moulin:1996vv,Moulin:1999hwj,Bernard:2000ovj}.
However, recent advances in high-intensity laser technology have substantiated the perspectives of a first measurement of this effect with state-of-the-art technology in the near future \cite{DiPiazza:2011tq,Battesti:2012hf,King:2015tba,Karbstein:2019oej,Fedotov:2022ely}. 
One of the key challenges is to achieve a sufficiently large signal-to-noise ratio allowing to measure the small quantum vacuum signal in the presence of the huge number of photons constituting the driving laser fields.

In the present work we put forward a specific setup allowing for its detection in an experiment based on the collision of two focused high-intensity laser pulses in a counter-propagating geometry; cf. also Refs.~\cite{Aleksandrov:1985,Tommasini:2010fb,Monden:2012,King:2012aw,Dinu:2013gaa,Karbstein:2014fva,Ataman:2018ucl,Pegoraro:2021whz,Roso:2021hfo,Ataman:2023gin}.
For recent measurements of signatures of light-by-light scattering processes using the strong Coulomb fields of heavy ions, see Refs. \cite{ATLAS:2017fur,CMS:2018erd,ATLAS:2019azn,Brandenburg:2022tna}.
The basic idea of our setup is to use the same focusing optics for both pulses to be collided and to employ a dark-field approach for the detection of the induced single-photon-level signal.
To this end two subsequent pulses generated by the same laser system and appropriately separated in time by means of a delay line are collided in their common focus after the propagation direction of the first pulse has been reversed at a spherical retro-reflector.
Moreover, a central shadow is imprinted in the transverse profile of the initial beam by reflecting the initial beam off a mirror with hole. By construction this shadow is then present both in the converging beam prior to its focus and the diverging beam after its focus, while a peaked on-axis focus profile is retained \cite{Karbstein:2020gzg,Karbstein:2022uwf}. The shadow in the diverging beam is effectively imaged onto a single-photon sensitive detector via a hole in the retro-reflector such as to spatially filter out a sizable fraction of the quantum vacuum signal induced in the collision of the two laser pulses.

Our article is organized as follows:
In Sec.~\ref{sec:form} we explain the scenario put forward for the measurement of photonic  quantum vacuum signals in the present work. To this end, we briefly recall how the relevant quantum vacuum signal can be derived from \Eqref{eq:HElagrangian}. Our method of choice is the {\it vacuum emission picture} \cite{Galtsov:1971xm,Karbstein:2014fva} for which a flexible and convenient numerical solver allowing to study all-optical signatures of quantum vacuum nonlinearities in generic laser fields is available \cite{Blinne:2018nbd}.
Thereafter we explain the experimental setup devised by us in full detail.
Finally, we discuss the theoretical modeling of the driving laser fields and comment on the numerical implementation of the considered scenario in the vacuum emission solver \cite{Blinne:2018nbd} which we employ to evaluate the signal photon yield and emission characteristics for our setup.
In Sec.~\ref{sec:results} we present results for the prospective quantum vacuum signals attainable in our setup using state-of-the-art laser parameters as input.
Finally, we close with conclusions and a brief outlook in Sec.~\ref{sec:concls}.

\section{Scenario}\label{sec:form}

\subsection{Theoretical Basics}\label{subsec:theory}

Photonic quantum vacuum signals can be conveniently analyzed by viewing them as vacuum emission processes stimulated by the applied macroscopic electromagnetic field $F^{\mu\nu}$ \cite{Galtsov:1971xm,Karbstein:2014fva}. 
In the parameter regime considered throughout this work, single signal photon emission vastly dominates over emission processes with higher multiplicities.
The amplitude for the relevant zero-to-single signal photon transition can be compactly expressed as \cite{Karbstein:2014fva}
\begin{equation}
    {\cal S}_p(\vec{k}) = \bigl\langle\gamma_p(\vec{k})\bigr|\int{\rm d}^4x\,\frac{\partial{\cal L}_{\rm int}}{\partial F^{\mu\nu}}(x)\hat{f}^{\mu\nu}(x)\bigl|0\bigr\rangle
    =-2{\rm i}k^\alpha\frac{\epsilon_{(p)}^{*\mu}(\vec{k})}{\sqrt{2k^0}}\int{\rm d}^4x\,{\rm e}^{-{\rm i}kx}\,\frac{\partial{\cal L}_{\rm int}}{\partial F^{\alpha\mu}}(x)\,\Bigg|_{k^0=|\vec{k}|}\,,
\label{eq:amplitude}
\end{equation}
where the in-state $|0\rangle$ denotes the vacuum subjected to the initially applied field featuring zero signal photons by definition, and $\langle\gamma_p(\vec{k})\bigr|$ is the out-state containing a single on-shell transverse signal photon of wave vector $\vec{k}$ and polarization $p$; polarization vector $\epsilon_{(p)}^\mu(\vec{k})$.
Finally, $\hat{f}^{\mu\nu}(x)$ denotes the canonically quantized field strength tensor of the signal photon field.
Equations~\eqref{eq:HElagrangian} and \eqref{eq:amplitude} imply that determining the signal emission amplitude effectively boils down to performing a four-dimensional Fourier integral of a function cubic in $F^{\mu\nu}$.

The signal-photon polarizations can be parameterized as $\epsilon_{(p)}^\mu(\vec{k})=\bigl(0,\vec{e}_p(\vec{k})\bigr)$ by two orthonormal three vectors $\vec{e}_p(\vec{k})$ fulfilling $\vec{k}\cdot\vec{e}_p(\vec{k})=0$.
For a linear polarization basis we have $p\in\{1,2\}$. The associated vectors $\vec{e}_1(\vec{k})$, $\vec{e}_2(\vec{k})$ are real and fulfill $\vec{k}/|\vec{k}|\times\vec{e}_1(\vec{k})=\vec{e}_2(\vec{k})$.
On the other hand, for a circular polarization basis we have $p\in\{+,-\}$ and $\vec{e}_\pm=\bigl\{\vec{e}_1(\vec{k})\pm{\rm i}\vec{e}_2(\vec{k})\bigr\}/\sqrt{2}$. Here, ``$+$" denotes right and ``$-$" left hand circular polarization, respectively.
The above definitions imply $\epsilon_{(p)\mu}(\vec{k})\epsilon_{(p')}^{*\mu}(\vec{k})=\delta_{p,p'}$ for both cases.
We also note that generic elliptic polarizations can be parameterized by polarization vectors $\vec{e}_\epsilon(\vec{k})=\{c_1(\vec{k})\,\vec{e}_1(\vec{k})+{\rm i}\,c_2(\vec{k})\,\vec{e}_2(\vec{k})\}/\sqrt{c_1^2(\vec{k})+c_2^2(\vec{k})}$, with real-valued functions $c_1(\vec{k})$, $c_2(\vec{k})$.

The differential number of signal photons of energy ${\rm k}\equiv|\vec{k}|$, propagation direction $\vec{k}/|\vec{k}|$ and polarization $p$ associated with the transition amplitude~\eqref{eq:amplitude} can then be expressed as \cite{Karbstein:2014fva}
\begin{equation} \label{eq:diffphotonnumber}
    {\rm d}^3N_p(\vec{k})=\frac{{\rm d}^3k}{(2\pi)^3}\,\bigl|{\cal S}_p(\vec{k})\bigr|^2\,.
\end{equation}
Clearly, the signal attainable in a polarization insensitive measurement follows upon summation over two transverse signal polarizations as ${\rm d}^3 N(\vec{k})= \sum_{p} {\rm d}^3 N_p(\vec{k})$.
For the analysis of the signal in \Eqref{eq:diffphotonnumber} it is particularly convenient to employ spherical momentum coordinates where ${\rm d}^3k={\rm k}^2\,{\rm dk}\,{\rm d}\varphi\,{\rm d}\!\cos\vartheta$.
The angular emission characteristics of the signal is then encoded in
\begin{equation} \label{eq:photonnumberdens}
    \frac{{\rm d}^2N_p(\varphi,\vartheta)}{{\rm d}\varphi\,{\rm d}\!\cos\vartheta}=\frac{1}{(2\pi)^3}\int_0^\infty{\rm dk}\,{\rm k}^2\,\bigl|{\cal S}_p(\vec{k})\bigr|^2\,.
\end{equation}
In the present work we will ultimately employ the numerical solver put forward in Ref.~\cite{Blinne:2018nbd} to evaluate the signals in  Eqs.~\eqref{eq:diffphotonnumber}
and \eqref{eq:photonnumberdens} for the specific laser field configuration to be implemented in our experimental setup detailed in Sec.~\ref{sec:expsetup}.

In the parameter regime where the driving laser fields can be accurately modeled as leading-order paraxial beams detailed analytical insights into \Eqref{eq:diffphotonnumber} are possible.
This is especially true for the collision of just two beams: here, the signal decomposes in two distinct main contributions that can be understood in terms of laser photons of one beam (``the probe") being quasi-elastically scattered off the cycle-averaged intensity profile of the other beam (``the pump") and vice versa \cite{Karbstein:2016lby}.
Due to the nonlinear interaction of the driving laser fields, in this case the signal generically comprises components polarized similar ($\parallel$) and perpendicular ($\perp$) to the laser photons constituting the probe field.
It can be shown that in the low-energy limit considered throughout this work the associated signal photon numbers scale as
\begin{equation}
 N_{\parallel,\perp}\sim c_{\parallel,\perp}\,\alpha^4 \Bigl(\frac{W_{\rm pump}}{m}\frac{\omega_{\rm probe}}{m}\Bigr)^2 N_{\rm probe}\,,
 \label{eq:scaling}
\end{equation}
with $\alpha$, the pulse energy of the pump $W_{\rm pump}$, the probe photon energy $\omega_{\rm probe}$ and the probe photon number $N_{\rm probe}\simeq W_{\rm probe}/\omega_{\rm probe}$. 
Finally, the coefficients $c_{\parallel,\perp}$ governing the strength of the signal components induced in a $\parallel,\perp$-polarized mode depend on both the low-energy constants~\eqref{eq:a,b} and the polarizations of the driving laser fields.  
Besides, the signal photon numbers typically depend non-trivially on other characteristic parameters of the colliding beams such as the collision angle, their pulse durations, waist sizes and Rayleigh ranges; cf., e.g., Ref.~\cite{Mosman:2021vua,Karbstein:2021ldz}. Specifically, the dependence on the collision angle is such that the signal vanishes for co-propagating beams and becomes maximum for counter-propagation.
In the paraxial limit, these additional dependencies are encoded in a function that amounts to global factor in \Eqref{eq:scaling} in the sense that it is independent of the polarization characteristics of the driving beams as well as the signal.
In turn, ratios of the explicit values of $c_{\parallel,\perp}$ provided below reproduce the ratios of the associated signal photon numbers.
Note, however, that higher-order corrections in the diffraction angles characterizing both the colliding beams and the signal, which can for instance be safely neglected for state-of-the-art XFEL probes, generically break this factorization. Besides, beyond the low-energy limit also higher-order derivative corrections \cite{Gusynin:1995bc,Gusynin:1998bt,Karbstein:2021obd} to \Eqref{eq:HElagrangian} are expected to impact the signal polarizations \cite{Dinu:2013gaa,King:2023eeo}.
Because in the present context we are only interested in an estimate for the signals to provide for a reference and guidance for our full numerical calculations detailed below, in our analytical calculations we restrict ourselves to the paraxial limit. 
In this case, the coefficients $c_{\parallel,\perp}$ in \Eqref{eq:scaling} are given by
\begin{align}
    c_\parallel&\simeq([a+b)+(a-b)\cos(2\phi)]^2\,, \nonumber\\
    c_\perp&\simeq[(a-b)\sin(2\phi)]^2\,,
    \label{eq:c_lin}
\end{align}
for linearly polarized beams; $\phi$ measures the relative polarization of the colliding beams.
On the other hand, if both beams are circularly polarized we have
\begin{align}
    c_\parallel&\simeq(a+b)^2\,, \nonumber\\
    c_\perp&\simeq0\,,
    \label{eq:c_circ}
\end{align}
independent of the helicities of pump and probe.
The same result~\eqref{eq:c_circ} is obtained for a linearly polarized probe being collided with a circularly polarized pump. 
Finally, for a circularly polarized probe and a linearly polarized pump we find
\begin{align}
    c_\parallel&\simeq(a+b)^2\,, \nonumber\\
    c_\perp&\simeq(a-b)^2\,.
    \label{eq:c_circlin}
\end{align}
These results for $c_{\parallel,\perp}$ clearly depend on the low-energy constants $a$ and $b$ only in terms of $a+b$ and $a-b$.
The numbers of signal photons $N$ attainable in a polarization insensitive measurement follow by adding the $\parallel$ and $\perp$ polarized signals, i.e., are given by \Eqref{eq:scaling} with $c_{\parallel,\perp}\to c_\parallel+c_\perp$.
We will make use of these dependencies to benchmark our numerical results in Sec.~\ref{sec:results}.

A comparison of the results in Eqs.~\eqref{eq:c_lin}-\eqref{eq:c_circlin} then implies that the maximum signal photon number is obtained for linearly polarized pump and probe beams and $\phi=\pi/2$ \cite{Karbstein:2019bhp} for which $c_\parallel+c_\perp\simeq4b^2$.
Because $c_\perp\simeq0$ in this case, this number also amounts to the maximum possible signal polarized similarly to the probe.
On the other hand, the maximum polarization/helicity-flip signal is obtained in the collision of linearly polarized pump and probe beams with $\phi=\pi/4$ \cite{Aleksandrov:1985} or a circularly polarized probe collided with a linearly polarized pump, respectively. Both of these cases result in $c_\perp=(a-b)^2$.

Also note that Eqs.~\eqref{eq:c_lin} and \eqref{eq:c_circlin} imply that when colliding either linearly or circularly polarized probe light with a linearly polarized pump and measuring both signal components -- at least in principle -- the values of $a$ and $b$ can be inferred separately \cite{Karbstein:2022uwf}.

\subsection{Experimental Setup}\label{sec:expsetup}

\begin{figure}
  \centering
  \includegraphics[width=0.9\linewidth]{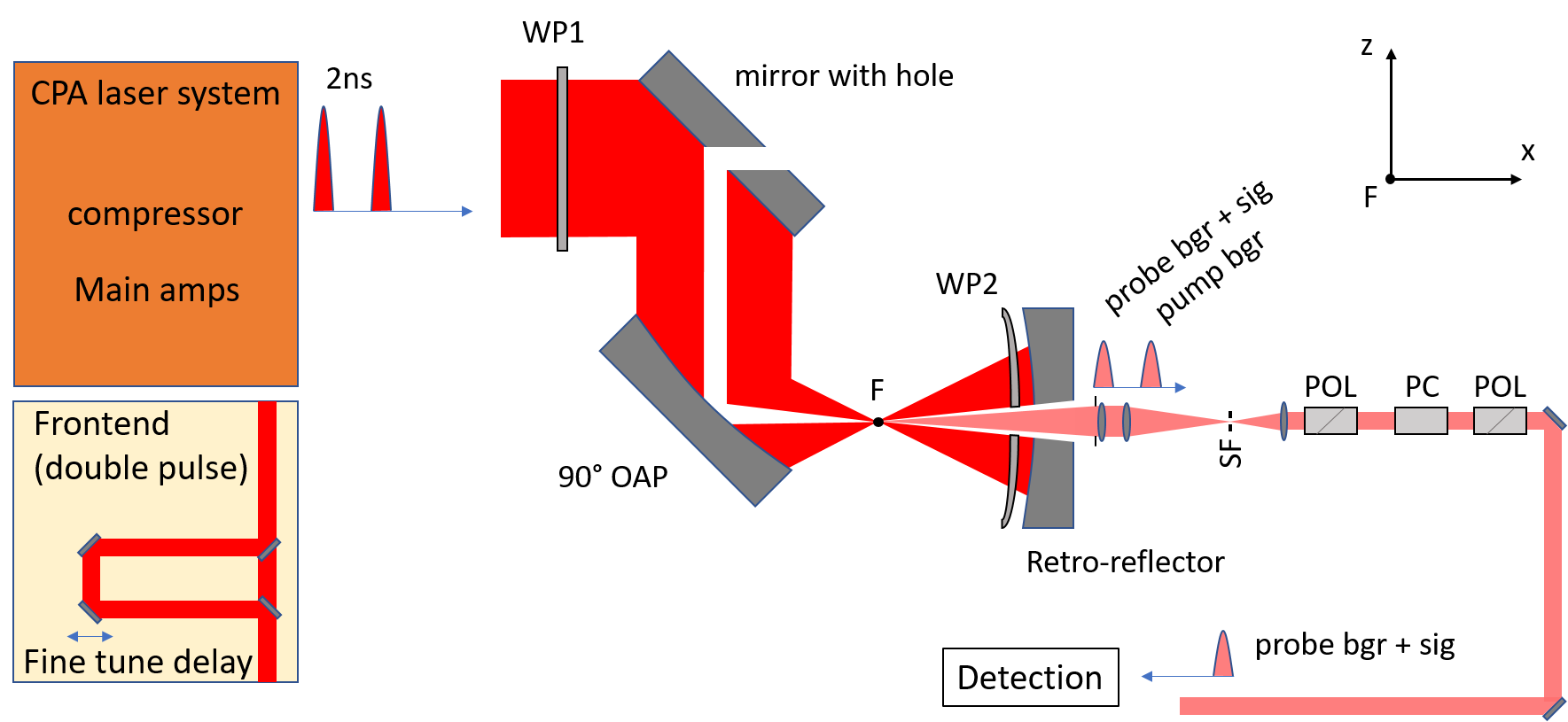}
  \caption{Schematic layout of the experimental setup. Two pulses generated by the same frontend are separated in time and fed into the same focusing optics such that they are collided in their common focus after the propagation direction of the first pulse has been reversed at a retro-reflector.
  A central shadow is imprinted in the transverse profile of the initial beam by reflecting it off a mirror with hole. The shadow is effectively imaged onto a single-photon sensitive detector via another hole in the retro-reflector such as to spatially filter out a sizable fraction of the quantum vacuum signal while minimizing the background.
  The remaining pump background traversing the hole in the retro-reflector about $2\,{\rm ns}$ in advance to the signal can be filtered out by a Pockels cell (PC). 
  Abbreviations: Wave plate (WP), off-axis parabolic mirror (OAP), focal point (F), spatial filter (SF), polarizer (POL).}
  \label{fig:exp_setup}
\end{figure}

The experimental setup envisioned by us to allow for the measurement of quantum vacuum signals in an all-optical experiment at a petawatt (PW) class laser system, such as the $3\,{\rm PW}$ Advanced Titanium-Sapphire Laser System (ATLAS$\,3000$) at the Centre for Advanced Laser Applications (CALA) \cite{CALA} in Garching, Germany, is depicted schematically in Fig.~\ref{fig:exp_setup}.
We have decided for a counter-propagating geometry of the driving laser fields to maximize the signal-photon yield in \Eqref{eq:scaling}.
With regard to an experimental implementation in the high-intensity domain, this geometry has so far been excluded for optical frequencies because the signal photons predominantly emerge in the forward cones of the driving laser beams and are thus completely background dominated.
In addition, the large diameter of the amplified high-intensity beam, which is typically of the order of several tens of centimeters, and the strong focusing required for a significant signal yield pose challenges on possible schemes even with only two colliding pulses.

We address this challenge by generating a double pulse each containing an energy of several tens of Joules. The two pulses will be introduced in the frontend of the high-intensity laser system by splitting the original pulse, introducing an appropriate temporal delay of about $2\,{\rm ns}$. The delay and the pulse energy content can be finely tuned by a delay stage and an appropriate continuous Neutral Density (ND) filter. Subsequently these pulses will travel the exact same path through the amplifiers, compressor and the laser beam delivery to the experimental
vacuum chamber, where a $90^\circ$ off-axis parabolic mirror (OAP) with effective focal length of $f_{\rm eff} = 30\,{\rm cm}$ focuses the pulses. A spherical mirror with suitable radius of curvature ($R = 30\,{\rm cm}$) will be mounted downstream of the primary focus point. It will reflect and refocus the pulses, such that the first pulse on its return collides with the second pulse in its first pass in the common focus spot.
A motorized hexapod will enable precise spatial overlap, whereas fine tuning the temporal delay between the pulses in the frontend varies temporal overlap.
The spatial overlap between the pulses can be optimized to better than $<0.1\,\upmu{\rm m}$ with typically available hardware.
At the same time, fine tuning the temporal delay between the pulses in the frontend to vary the temporal overlap is possible to better than ${\rm fs}$ precision.
We note, however, that this overlap will be influenced by shot-to-shot variations, especially of the beam pointing, present in any large laser system. In principle, these can be minimized by active control to a similar or better level.
The necessary level of improvement will depend on the real-world background contributions and thus is beyond the scope of this manuscript.

By introducing two large aperture wave plates, one before the OAP (WP1) and one in front of the spherical retro-reflecting mirror (WP2), we can change the polarization of both pump and probe pulses and thus realize all potentially relevant polarization combinations studied in Sec.~\ref{subsec:theory} in experiment; see Fig.~\ref{fig:exp_setup}. To this end, we consider all incident laser pulses to be linearly polarized.
In turn, if WP1 is a quarter wave plate and WP2 is an eighth wave plate, the probe will be circularly polarized and the pump polarization will be linear. If WP1 is omitted and only an eighth wave plate is used for WP2, the probe will be linearly polarized and the pump circularly polarized. On the other hand, if we use a quarter wave plate for WP1 and omit WP2, probe and pump will be circularly polarized with opposite helicity. If both WP1 and WP2 are quarter wave plates, probe and pump will also be circularly polarized but with equal helicity.
Finally, if we do not use WP1 and operate WP2 as a quarter wave plate we can rotate the polarization of the pump pulse by an angle of $0\leq \phi \leq 90^{\circ}$ relative to that of the probe.
This allows us to tune the coefficients in \Eqref{eq:c_lin}. In particular, recall that $\phi=45^\circ$ maximizes the polarization-flip signal $N_\perp$. Even though being reduced by a factor of about $c_\perp/(c_\parallel+c_\perp)\big|_{\phi=45^\circ}\simeq0.07$ relative to the total signal, this signal can likely be more efficiently separated from the residual co-propagating laser background that is inevitable in any actual experimental implementations of our setup using polarimetry.

The second key-challenge, namely achieving a sufficient signal-to-background separation in experiment, is mastered by an approach pioneered by Refs.~\cite{Peatross:1994,Zepf:1998} for the detection of weak nonlinear optics signals in the presence of strong fields.
Its great potential for the measurement of quantum vacuum signals has recently been showcased in Refs.~\cite{Karbstein:2020gzg, Karbstein:2022uwf} for the case of high-intensity and x-ray free electron laser (XFEL) collisions.
The idea underlying this approach is to remove a central part of the collimated beam by reflecting the initial beam at a mirror with hole to create an annular beam.
Alternatively the annular beam profile can be introduced by an obstacle put into the beamline.

The central shadow implemented in the transverse beam profile is then present in both the convergent beams prior to focusing and the divergent beams after focusing; see Fig.~\ref{fig:exp_setup}.
At the same time, a strongly peaked on-axis focus profile, very similar to that of an ordinary focused beam without dark field, is retained; in the focus the information about the central shadow is encoded in the Airy ring structure \cite{Karbstein:2020gzg}.
Because the peak-field driven quantum vacuum signal is predominantly induced by the central focus peak, which -- to leading order -- can locally not be distinguished from the focus profile of a Gaussian beam, a sizable signal fraction is thus expected to be emitted into the dark field. Our setup in Fig.~\ref{fig:exp_setup} is designed such as to image the signal induced in forward direction via a hole in the retro-reflector onto a single-photon sensitive detector, and thus to allow for a spatial separation of the quantum vacuum signal from the background of the driving laser photons.
To reduce background due to diffracted photons in this shadow region, a beam block will be inserted closer to the primary focus. A set of two lenses positioned behind the hole in the retro-reflector images the beam block onto the entrance aperture of a spatial filter (SF) with pinhole matched to the image of the high-intensity focal spot. This arrangement prevents singly scattered photons from the OAP from propagating to the detector and we hence have a situation where only doubly scattered photons could enter the beam path. The photons that pass through the pinhole of the spatial filter naturally decompose into (i) a pure background component due to the pump pulse and (ii) a probe background plus signal component arriving delayed by $\Delta t=2R/c$.
Ideally, the background contribution is negligible with respect to the signal.

In Refs.~\cite{Karbstein:2020gzg, Karbstein:2022uwf} a relation between the inner $\theta_{\rm in}$ and outer $\theta_{\rm out}$ radial divergences of a paraxial beam of wavelength $\lambda$ featuring a circular flat top near-field profile with a perfect central circular shadow and its $1/{\rm e}^2$ beam waist $w_0$ (w.r.t. intensity) was derived. This can be expressed as
\begin{equation}
    w_0\simeq\frac{\lambda}{\pi\theta_{\rm out}}\sqrt{\frac{2(1-{\rm e}^{-1})}{1+\nu}}\,,\quad\text{where}\quad \nu=\Bigl(\frac{\theta_{\rm in}}{\theta_{\rm out}}\Bigr)^2
    \label{eq:w0parax}
\end{equation}
measures the fraction of the transverse area of the beam in the near field obstructed by the central shadow; in the limit of $\nu\to0$ we recover the result for a flat top without shadow.
Assuming the full beam entering our setup in Fig.~\ref{fig:exp_setup} to feature a flat top transverse profile into which the central shadow is imprinted and neglecting the impact of the asymmetry induced by the $90^\circ$ OAP in our setup for the moment, the parameter $\nu$ can also be related to the beam energies of the initial beam $W_0$ and the resulting annular beam as $W=W_0(1-\nu)$.
Clearly, $\theta_{\rm out}$ is to be kept fixed and amounts to the radial opening angle of the focused beam. At this point we also emphasize that in the present context the paraxial approximation should allow for reliable insights as long as $\theta_{\rm out }\ll1$.
Moreover, it can be shown that the focus peak field amplitude of such a beam is given by \cite{Karbstein:2023}
\begin{equation}
 {\cal E}_{\rm peak}\simeq(1-\nu)\,\sqrt{\frac{1-{\rm e}^{-1}}{1+\nu}}\,{\cal E}_0\,,
 \label{eq:E0parax}
\end{equation}
where ${\cal E}_0$ is the peak field amplitude of a Gaussian beam of energy $W_0$ featuring the same waist size and temporal envelope.
Also note that by construction in our setup the pump and probe laser pulses have the same pulse duration $\tau$ and the same waist $w_0$ when they collide in their common focus.

Using the above findings in the analytical approximation for the angular emission characteristics in the collision of two pulsed Gaussian beams provided in Eq.~(11) of Ref.~\cite{Mosman:2021vua} and integrating over a forward cone with radial opening angle $\theta_{\rm det}\leq\theta_{\rm in}$ we thus can arrive at an estimate for the quantum vacuum signal to be measured by a detector with an angular acceptance of $\pi\theta_{\rm det}^2$ located behind the central hole in the retro-reflector in Fig.~\ref{fig:exp_setup}.
For completeness, we emphasize that though formally resorting to an infinite Rayleigh range approximation for the probe, the analytical approximation~\cite{Mosman:2021vua} should allow for reasonable estimates even for the collision of beams with the same Rayleigh range as considered here: the somewhat less pronounced localization of the composite laser field in the interaction region in longitudinal direction and the stronger localization in transverse direction associated with this idealization tend to compensate each other.

Identifying the beam waists with \Eqref{eq:w0parax}, equating the pulse durations with $\tau$ and multiplying by an overall rescaling factor of $({\cal E}_{\rm peak}/{\cal E}_0)^6$ to account for the reduction of the peak field amplitude, Eq.~(11) of Ref.~\cite{Mosman:2021vua} provides us with an estimate for the number of $p\in\{\parallel,\perp\}$ polarized signal photons  reaching the detector,
\begin{align}
 N_{p,{\rm det}}&\simeq c_p\,\sqrt{\frac{3}{\pi}}\,\frac{2(1-{\rm e}^{-1})}{6075}\,(\pi\alpha)^4\,\Bigl(\frac{W_0}{m}\Bigr)^3\Bigl(\frac{\lambdabar_{\rm C}}{\lambda}\Bigr)^5\,\theta_{\rm out}^4\,\frac{(1-\nu)^6}{1+\nu}\,\sqrt{F_0F_1} \nonumber\\
 &\quad\quad\quad\times\biggl(1-\exp\biggl\{-\frac{4}{3}\frac{1-{\rm e}^{-1}}{1+\nu}\sqrt{\frac{F_1}{F_0}}\Bigl(\frac{\theta_{\rm det}}{\theta_{\rm out}}\Bigr)^2\biggr\}\biggr) \,,
 \label{eq:Npapprox}
\end{align}
where 
\begin{align}
F_\beta&:=F\biggl(\frac{8}{\pi}\sqrt{\frac{2}{3}}\,\frac{1-{\rm e}^{-1}}{1+\nu}\,\frac{\lambda}{\tau}\,\frac{\sqrt{1+2\beta^2}}{\theta_{\rm out}^2}\biggr)\,,\quad\text{with} \nonumber\\
F(\chi)&:=\chi^2\,{\rm e}^{2\chi^2}\int_{-\infty}^\infty{\rm d}\kappa\,{\rm e}^{-\kappa^2}\Bigl[{\rm e}^{2\kappa\chi}\,{\rm erfc}(\chi+\kappa)+{\rm e}^{-2\kappa\chi}\,{\rm erfc}(\chi-\kappa)\Bigr]^2\,,
\end{align}
and $\tau$ denotes the $1/{\rm e}^2$ pulse duration w.r.t. intensity.
In the limit where $\theta_{\rm det}\to \theta_{\rm in}$ and the maximum possible number of signal photons is detected within the hole in the retro-reflector we clearly have $(\theta_{\rm det}/\theta_{\rm out})^2\to\nu$.
Adopting this choice, for a wavelength of $\lambda=800\,{\rm nm}$, a typical pulse duration of $\tau_{\rm FWHM}=30\,{\rm fs}$ ($\tau\simeq51\,{\rm fs}$) available at state-of-the-art high-intensity laser systems and an outer radial beam divergence as large as $\theta_{\rm out}\simeq1/2$ such as to reach a large peak field strength in the focus, \Eqref{eq:Npapprox} allows us to infer that the maximum signal can be achieved for $\nu\simeq0.1$.
We emphasize that, because $\theta_{\rm out}\simeq1/2$ does not fulfill the citerion $\theta_{\rm out}\ll1$, higher-order corrections to the (leading-order) paraxial approximation adopted in the derivation of Eqs.~\eqref{eq:w0parax}-\eqref{eq:Npapprox} may become sizable in the considered limit. Especially as it makes use of an infinite Rayleigh range approximation, \Eqref{eq:Npapprox} thus is likely to allow only for a qualitative estimate for the present parameters.

When aiming at implementing the experimental setup outlined above at the ATLAS$\,3000$ system where the beam radius is $r_{\rm beam}=14\,{\rm cm}$, we obtain $\theta_{\rm out}\approx r_{\rm beam}/f_{\rm eff}\simeq0.47$.
Correspondingly, because of $\nu\simeq(r_{\rm hole}/r_{\rm beam})^2$ the optimal value for the radius of the hole in the first mirror and the retro-reflector should be of the order of $r_{\rm hole}\approx\sqrt{0.1}\,r_{\rm beam}\approx4.5\,{\rm cm}$.
Specializing to ATLAS$\,3000$ parameters in the remainder of this work, we choose a somewhat larger value of $r_{\rm hole}=4.75\,{\rm cm}$ for the hole in the first mirror and $r_{\rm det}=3.75\,{\rm cm}$ for the one in the retro-reflector that effectively defines the size of the detection region. Correspondingly, we have $\nu\simeq0.12$.
These hole sizes are manageable in terms of available, high quality optics for collection and subsequent signal filtering and analysis. 
The experimental results in Ref.~\cite{Doyle:2021mdt} suggest that double scattering at the OAP and in the collection lens will contribute the majority of background photons within a gate window width of $1\,{\rm ns}$ around the laser pulse. Importantly, these are significantly delayed or advanced with respect to the signal photons and can hence be further suppressed by appropriate temporal gating techniques.

\subsection{Laser Field Model}\label{sec:lasermodel}

As detailed above, the $90^\circ$ OAP constitutes a key component of our setup.
For the modeling of its impact on the driving laser fields we resort to the vector diffraction formulae allowing to infer the electromagnetic field near the focus of an off-axis paraboloid for a given incident field configuration derived in Ref.~\cite{Bahk:2005} on the basis of the Stratton–Chu formula \cite{Stratton:1939}. See also Ref.~\cite{Zeng:2019} for a recent systematic investigation. 
This in particular ensures that the distortions of the laser fields when reflected off the OAP are consistently accounted for in our calculation of the quantum vacuum signals.

The input to this calculation is the transverse profile of a monochromatic paraxial beam specified at a longitudinal coordinate ${\rm z}={\rm z_0}$ right in front of the OAP such that diffraction effects from the input plane to the OAP surface can be safely neglected.
This beam propagates in negative $\rm z$ direction and is characterized by a complex electric field of the form $\vec{E}_{\rm in}(x)={\rm e}^{-{\rm i}\omega t}\,\vec{E}_{\rm in}(\omega,\vec{x})$ with
\begin{equation}\label{eq:incidentfield}
    \vec{E}_{\rm in}(\omega,\vec{x})\big|_{{\rm z}={\rm z}_0} = {\rm e}^{-{\rm i}\omega {\rm z}_0} \bigl[E_{0,{\rm x}}({\rm x},{\rm y})\,\vec{e}_{\rm x} + E_{0,{\rm y}}({\rm x},{\rm y})\,\vec{e}_{\rm y}\bigr]\,,
\end{equation}
and a corresponding magnetic field given by $\vec{B}_{\rm in}=-\vec{e}_{\rm z}\times\vec{E}_{\rm in}$.
Here, $\omega$ denotes the oscillation frequency of the beam and the transverse profile functions $E_{0,{\rm x}}({\rm x},{\rm y})$ and $E_{0,{\rm y}}({\rm x},{\rm y})$ implement transverse beam sizes much larger than its wavelength $\lambda=2\pi/\omega$ \cite{Bahk:2005}.
The origin of our coordinate system is in the focal point F in Fig.~\ref{fig:exp_setup}. For a monochromatic input beam, the beam reflected off the OAP is, of course, also monochromatic.
For a $90^\circ$ OAP the electric field of the resulting beam near its focus at $\vec{x}=0$ (see below for the precise condition) can then be expressed as \cite{Bahk:2005}
\begin{align}
    E_{\rm x}(\omega,\vec{x}) &= \frac{{\rm i}\omega}{2\pi}\iint {\rm d}\tilde{\rm x} \, {\rm d}\tilde{\rm y} \biggl[ E_{0,{\rm x}}\biggl(1-\frac{\tilde{\rm x}^2}{2f^2(1+s)}\biggr) 
    - E_{0,{\rm y}}\, \frac{\tilde{\rm x}\tilde{y}}{2f^2(1+s)}\biggr]\frac{1}{f(1+s)}\, {\rm e}^{-{\rm i}\omega\phi}\,, \nonumber\\
    E_{\rm y}(\omega,\vec{x}) &= \frac{{\rm i}\omega}{2\pi}\iint {\rm d}\tilde{\rm x} \, {\rm d}\tilde{\rm y} \biggl[E_{0,{\rm y}}\biggl(1-\frac{\tilde{\rm y}^2}{2f^2(1+s)}\biggr)-E_{0,{\rm x}} \, \frac{\tilde{\rm x}\tilde{\rm y}}{2f^2(1+s)}\biggr]\frac{1}{f(1+s)}\, {\rm e}^{-{\rm i}\omega\phi}\,, \nonumber\\
    E_{\rm z}(\omega,\vec{x}) &= \frac{{\rm i}\omega}{2\pi}\iint {\rm d}\tilde{\rm x} \, {\rm d}\tilde{\rm y}\, \bigl(E_{0,{\rm x}}\tilde{\rm x} + E_{0,{\rm y}}\tilde{\rm y}\bigr) \frac{1}{f^2(1+s)^2}\, {\rm e}^{-{\rm i}\omega\phi}\,,
    \label{eq:E_OAP}
\end{align}
with the phase
\begin{equation}
    \phi = \frac{\tilde{\rm x}{\rm x} + \tilde{\rm y}{\rm y}+f(s-1){\rm z}}{f(1+s)}
    \label{eq:phi}
\end{equation}
containing the full dependence of \Eqref{eq:E_OAP} on $\vec{x}$. Here, $f=f_{\rm eff}/2$ is the parent focal length of the $90^\circ$ OAP, $E_{0,{\rm x}}\equiv E_{0,{\rm x}}(\tilde{\rm x},\tilde{\rm y})$, $E_{0,{\rm y}}\equiv E_{0,{\rm y}}(\tilde{\rm x},\tilde{\rm y})$, $s=(\tilde{\rm x}^2+\tilde{\rm y}^2)/f_{\rm eff}^2$ and the integrations are performed over the transverse coordinates $\tilde{\rm x}$ and $\tilde{\rm y}$ of the input beam. The associated magnetic field $\vec{B}(\omega,\vec{x})$ is fully determined by Maxwell's equations in vacuum \cite{Bahk:2005} and thus can be readily extracted from \Eqref{eq:E_OAP}; for its explicit expression see Ref.~\cite{Bahk:2005}.
The use of \Eqref{eq:E_OAP} is well-justified given the {\it near focus} condition $|\vec{x}|^2 \ll 2|\tilde{\rm x}{\rm x}+\tilde{\rm y}{\rm y}+f(s-1){\rm z}|$ is met \cite{Bahk:2005}, i.e., as long as $|\vec{x}|$ is sufficiently small and the integrals over $\tilde{\rm x}$ and $\tilde{\rm y}$ in \Eqref{eq:E_OAP} receive their main contributions from regions where this condition is met.

Also note that knowledge of the electromagnetic fields near the beam focus immediately grants access to the far-field angular distribution of the number ${\cal N}=W/\omega$ of laser photons per unit time. To this end, recall that in momentum space the (appropriately regularized) energy stored in the electromagnetic field can be expressed as
\begin{equation}
    W=\int{\rm d}^3k\,\frac{1}{2}\Bigl(\big|\vec{E}(\omega,\vec{k})\bigr|^2+\big|\vec{B}(\omega,\vec{k})\bigr|^2\Bigr)\,,
\end{equation}
which for a monochromatic beam with $\bigl|\vec{E}(\omega,\vec{k})\bigr|=\bigl|\vec{B}(\omega,\vec{k})\bigr|$ as considered here implies that
\begin{equation}
    \frac{{\rm d}^2{\cal N}(\varphi,\vartheta)}{{\rm d}\varphi\,{\rm d}\!\cos\vartheta}=\frac{1}{\omega}\int_0^\infty{\rm dk}\,{\rm k}^2\,\big|\vec{E}(\omega,\vec{k})\bigr|^2\,.
    \label{eq:dcalN}
\end{equation}
Making use of the fact that the electric field can be spanned by two orthogonal polarization vectors $\vec{e}_p(\vec{k})$ transverse to $\vec{k}$ (see Sec.~\ref{subsec:theory}), we can write
\begin{equation}
    \big|\vec{E}(\omega,\vec{k})\bigr|^2=\sum_p\big|\vec{e}_p(\vec{k})\cdot\vec{E}(\omega,\vec{k})\bigr|^2\,.
\end{equation}
This identity allows us to infer from \Eqref{eq:dcalN} that the far-field angular distribution of laser photons polarized in mode $p$ can be represented as (cf., e.g., Ref.~\cite{Karbstein:2019oej})
\begin{equation}
    \frac{{\rm d}^2{\cal N}_p(\varphi,\vartheta)}{{\rm d}\varphi\,{\rm d}\!\cos\vartheta}=\frac{1}{\omega}\int_0^\infty{\rm dk}\,{\rm k}^2\,\big|\vec{e}_p(\vec{k})\cdot\vec{E}(\omega,\vec{k})\bigr|^2\,.
    \label{eq:dcalNp}
\end{equation}

With \Eqref{eq:E_OAP}, and rewriting the relevant Fourier integral in the following form
\begin{align}
    \int{\rm d}^3x\,{\rm e}^{-{\rm i}(\vec{k}\cdot\vec{x}+\omega\phi)}
    =\Bigl(\frac{f_{\rm eff}}{\omega+k_{\rm z}}\Bigr)^2\,(2\pi)^3\,\delta\Bigl(\tilde{\rm x}+f_{\rm eff}\frac{k_{\rm x}
    }{\omega+k_{\rm z}}\Bigr)\,\delta\Bigl(\tilde{\rm y}+f_{\rm eff}\frac{k_{\rm y}}{\omega+k_{\rm z}}\Bigr)\,\delta({\rm k}-\omega)\,,
    \label{eq:Fourierint}
\end{align}
it is then straightforward to work out explicit expressions for Eqs.~\eqref{eq:dcalN} and \eqref{eq:dcalNp}. For completeness and future reference we provide the explicit results for the components of $\vec{E}(\omega,\vec{k})$ in Appendix~\ref{sec:E(omega,k)}.
In a coordinate system oriented such that $\vec{k}={\rm k}\,(\cos\vartheta,\sin\varphi\sin\vartheta,-\cos\varphi\sin\vartheta)$ especially the far-field angular distribution of the total number of laser photons~\eqref{eq:dcalN} can be expressed as
\begin{equation}
     \frac{{\rm d}^2{\cal N}(\varphi,\vartheta)}{{\rm d}\varphi\,{\rm d}\!\cos\vartheta}={\cal N}\,\frac{E^2_{0,{\rm x}}(\tilde{\rm x},\tilde{\rm y})+E^2_{0,{\rm y}}(\tilde{\rm x},\tilde{\rm y})}{(1-\cos\varphi\sin\vartheta)^2}\biggl(\int_0^{2\pi}{\rm d}\varphi\int_{-1}^1{\rm d}\!\cos\vartheta\,\frac{E^2_{0,{\rm x}}(\tilde{\rm x},\tilde{\rm y})+E^2_{0,{\rm y}}(\tilde{\rm x},\tilde{\rm y})}{(1-\cos\varphi\sin\vartheta)^2}\biggr)^{-1}\,,
     \label{eq:d2Nlaser}
\end{equation}
where
\begin{equation}
    \tilde{\rm x}=-f_{\rm eff}\,\frac{\cos\vartheta}{1-\cos\varphi\sin\vartheta}\quad\text{and}\quad\tilde{y}=-f_{\rm eff}\,\frac{\sin\varphi\sin\vartheta}{1-\cos\varphi\sin\vartheta}\,.
    \label{eq:tildexy}
\end{equation}
The only input needed to evaluate \Eqref{eq:d2Nlaser} is the transverse profile of the beam~\eqref{eq:incidentfield} impinging the $90^\circ$ OAP.
In this context, we also note that the ratio of \Eqref{eq:dcalNp} and $\cal N$ as determined from \Eqref{eq:dcalN} is manifestly finite and therefore specifically provides direct access to the fraction ${\cal N}_p/{\cal N}$ of $p$ polarized laser photons ${\cal N}_p$ contained in the total number of laser photons $\cal N$.

For a linearly polarized incident top-hat beam featuring a perfect central circular shadow we have
\begin{equation}
    E_{0,{\rm x}}({\rm x},{\rm y})=E_{0}({\rm x},{\rm y})\cos\beta\,,\quad E_{0,{\rm y}}({\rm x},{\rm y})=E_{0}({\rm x},{\rm y})\sin\beta\,,
  \label{eq:inputlinpol}
\end{equation}
where the angle $\beta$ parameterizes the possible polarization directions and the field profile reads
\begin{equation}\label{eq:incidentbeam}
    E_{0}({\rm x},{\rm y}) = \mathcal{E}_0\, \Theta(\sqrt{({\rm x}+f_{\rm eff})^2+{\rm y}^2}-r_{\rm hole})\,\Theta(r_{\rm beam} - \sqrt{({\rm x}+f_{\rm eff})^2+{\rm y}^2})\,,
\end{equation}
with Heaviside step function $\Theta(\cdot)$ and field amplitude ${\cal E}_0$, the value of which is left unspecified for the moment; see the discussion in the context of \Eqref{eq:E0} below.
On the other hand, for a ``$\pm$'' circularly polarized input beam we have
\begin{equation}
    E_{0,{\rm x}}({\rm x},{\rm y})=\frac{1}{\sqrt{2}}\,E_{0}({\rm x},{\rm y})\,,\quad E_{0,{\rm y}}({\rm x},{\rm y})=\mp{\rm i}\frac{1}{\sqrt{2}}\,E_{0}({\rm x},{\rm y})\,;
 \label{eq:inputcircpol}
\end{equation}
cf. also Sec.~\ref{subsec:theory}. To this end, recall that the beam impinging the $90^\circ$ OAP propagates in negative $\rm z$ direction.

In a next step, we plug either \Eqref{eq:inputlinpol} or \Eqref{eq:inputcircpol} into \Eqref{eq:incidentfield} and determine the resulting electric field~\eqref{eq:E_OAP} in the vicinity of the focus. The associated magnetic field is determined therefrom via Maxwell's equations.  
Accounting for the fact that the beam reflected off the $90^\circ$ OAP propagates along the positive $\rm x$ axis, we then multiply all field components with a factor of $\exp\bigl\{-({\rm x}-t)^2/(\tau/2)^2\bigr\}$ with $\omega\tau\gg1$ such as to implement a physical finite energy beam; $\tau$ is the $1/{\rm e}^2$ duration of the resulting pulsed beam w.r.t. intensity.
We denote the resulting (complex) electric and magnetic fields near focus by
\begin{equation}
    \vec{E}(x)={\rm e}^{-{\rm i}\omega t-\left(\frac{{\rm x}-t}{\tau/2}\right)^2}\vec{E}(\omega,\vec{x})\quad\text{and}\quad
    \vec{B}(x)={\rm e}^{-{\rm i}\omega t-\left(\frac{{\rm x}-t}{\tau/2}\right)^2}\vec{B}(\omega,\vec{x})\,.
    \label{eq:E(x),B(x)}
\end{equation}
The demand of $\omega\tau\gg1$ ensures that (i) the resulting pulsed beam has a negligible bandwidth and thus can be considered as quasi-monochromatic, and (ii) the violations of Maxwell's equations in vacuum inevitably introduced by this {\it ad hoc} prescription can be safely neglected; they are parametrically suppressed by inverse powers of $\omega\tau$.
For $\lambda=800\,{\rm nm}$ and $\tau_{\rm FWHM}=30\,{\rm fs}$ we have $\omega\tau\simeq314$ which implies that this assumption is indeed well-justified.
Modeling the electric field of the incident beam by the real part of \Eqref{eq:incidentfield} with transverse profile~\eqref{eq:incidentbeam} and adopting the above pulsed beam prescription, we then fix the field amplitude ${\cal E}_0$ by demanding the focus pulse energy $W=W_0\,(1-r^2_{\rm hole}/r^2_{\rm beam})$ to fulfill \cite{Blinne:2018nbd}
\begin{equation}
    W=\int{\rm d}t\int{\rm dy}\int{\rm dz}\,\bigl(\vec{e}_{\rm x}\cdot\vec{S}(x)\bigr)\big|_{{\rm x}=0}\,.
    \label{eq:E0}
\end{equation}
Here, $W_0$ is the total pulse energy of the employed laser system and $\vec{S}(x)={\rm Re}\{\vec{E}(x)\}\times{\rm Re}\{\vec{B}(x)\}$, with the electric and magnetic fields given in \Eqref{eq:E(x),B(x)}, is the Poynting vector.
ATLAS\,$3000$ can routinely provide pulses with an energy of $W_0=25\,{\rm J}$, which is the value adopted by us in the remainder of this work.
With the values for $r_{\rm beam}$ and $r_{\rm hole}$ envisioned for the implementation of our setup at ATLAS\,$3000$ given above we thus have $W\simeq22.1\,{\rm J}$.

\begin{figure}
  \centering
  \includegraphics[width=0.7\linewidth]{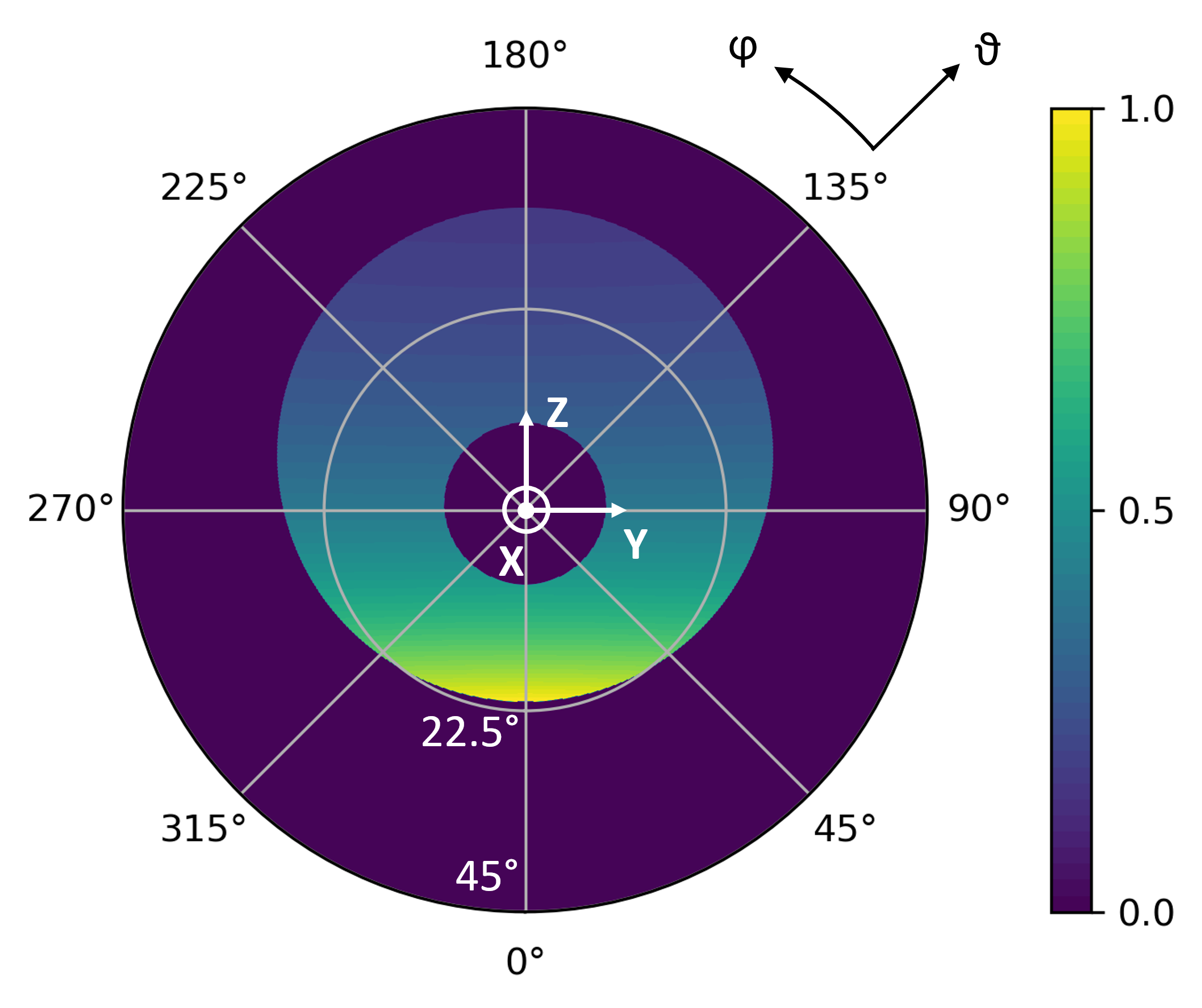}
  \caption{Normalized far-field angular distribution~\eqref{eq:d2Nlaser} of the laser photons constituting the probe beam in our setup; $f_{\rm eff}=30\,{\rm cm}$, $r_{\rm beam}=14\,{\rm cm}$, $r_{\rm hole}=4.75\,{\rm cm}$. The asymmetry is induced by the $90^\circ$ OAP.}
  \label{fig:probeanamoll}
\end{figure}

Figure~\ref{fig:probeanamoll} depicts the far-field angular distributions of the laser photons~\eqref{eq:d2Nlaser} of such a beam, namely the one constituting the probe beam in the experimental setup devised by us in Sec.~\ref{sec:expsetup} for the parameters available at ATLAS\,$3000$. This distribution features a pronounced asymmetry with regard to the $\rm xy$ plane, but is symmetric with respect to the $\rm xz$ plane.
We emphasize that because the distance from the focus F to the retro-reflector with hole is $R=30\,{\rm cm}$ in our setup and correspondingly $R/{\rm z}_{\rm R}\gg1$ assuming the detection to take place in the far field is well-justified;
recall that the Rayleigh range ${\rm z}_{\rm R}$ of the beam is the distance from the waist along its beam axis to the longitudinal coordinate where the  cross section area (characterized by its $1/{\rm e}^2$ average radius w.r.t. intensity) is doubled. To be specific, in our case we numerically infer ${\rm z}_{\rm R}\approx4.9\,\upmu{\rm m}$ and thus indeed have $R/{\rm z}_{\rm R}\approx6\times10^4\gg1$; the mean waist size is $w_0\approx708\,{\rm nm}$.

Finally, we assume the electric field of the laser beam reflected back to the focus by the retro-reflector to be again well-modelled by $\vec{E}(x)=\vec{E}(t,\vec{x})$ in \Eqref{eq:E(x),B(x)}. To be precise, as both the propagation direction and the orientation of the electric field of this beam are reversed at the mirror, the electric field of this beam follows by the mapping $\vec{E}(t,\vec{x})\to-\vec{E}(t,-\vec{x})$ and the associated magnetic field from Maxwell's equations. This choice ensures perfect collisions at zero impact parameter.
However, note that this construction does not require the colliding laser fields to have the same polarization. In fact, along the lines of Eqs.~\eqref{eq:inputlinpol} and \eqref{eq:inputcircpol} we can always choose a different polarization for the collimated incident top-hat beam impinging the OAP in our setup depicted in Fig.~\ref{fig:exp_setup} prior to implementing the above mapping.
We emphasize that throughout the present work we fix the polarization of a given beam in this way, which allows for an unambiguous, theoretically solid definition of the beam's polarization state.
In experiment, the polarization of the reflected beam can alternatively be modified with respect to the left-moving one by the appropriate insertion of optical elements such as half-wave and quarter-wave plates into its beam path; cf also Fig.~\ref{fig:exp_setup}.

\section{Results}\label{sec:results}

Instead of setting up a dedicated integrator allowing to directly evaluate the signal photon amplitude~\eqref{eq:amplitude} in the combined electromagnetic field of the left- and right-moving focused beams in our setup, we make use of the numerical vacuum emission solver \cite{Blinne:2018nbd} to determine the relevant quantum vacuum signals. This solver allows to initialize a given laser beam by providing its (complex) electric field at a fixed time discretized on a three dimensional grid. For a finite-energy beam, inherently coming with a spatial localization of the electromagnetic field, the input volume can -- at least in principle -- always be chosen large enough such as to accommodate the full laser field. In the present work, we identify the input time with the time where the maximum intensity is reached in the beam focus at $\vec{x}=0$. This amounts to setting $t=0$ in \Eqref{eq:E(x),B(x)} and thus using \begin{equation}
    \vec{E}(x)\big|_{t=0}={\rm e}^{-4\left(\frac{{\rm x}}{\tau}\right)^2}\vec{E}(\omega,\vec{x})\,,
\end{equation}
with the components of $\vec{E}(\omega,\vec{x})$ given in \Eqref{eq:E_OAP} to initialize a given laser beam in the vacuum emission solver \cite{Blinne:2018nbd}. The field configuration of the resulting beam is then self-consistently propagated to times $t\neq0$ by the solver according to Maxwell's equations in vacuum.
For further details on the numerical implementation see Appendix~\ref{sec:numimpl} and Ref.~\cite{Blinne:2018nbd}.

\subsection{Laser Polarizations}\label{subsec:laserpols}

Here, we provide results for seven different polarizations of the colliding laser beams. As clarified in the last paragraph of Sec.~\ref{sec:lasermodel}, in our theoretical considerations these are assumed to be implemented for the collimated beams prior to being reflected off the $90^\circ$ OAP in Fig.~\ref{fig:exp_setup}.
Namely, we consider (A) both beams to be linearly polarized with a relative polarization of
\begin{align}
    {\rm (A1):}&\quad\phi=0\phantom{/1}\quad(\beta_{\rm probe}=\beta_{\rm pump}=\pi/2)\,, \nonumber\\
    {\rm (A2):}&\quad\phi=\pi/4\quad(\beta_{\rm probe}=\pi/4\,,\ \beta_{\rm pump}=\pi/2)\,, \nonumber\\
    {\rm (A3):}&\quad\phi=\pi/2\quad(\beta_{\rm probe}=0\,,\ \beta_{\rm pump}=\pi/2)\,, \nonumber
\end{align}
where we also provide the explicit value of the angle $\beta$ used for the probe and pump laser fields in \Eqref{eq:inputlinpol} to realize a given relative polarization.
(B) Both beams are circularly polarized with the
\begin{align}
    {\rm (B1):}&\quad\text{opposite helicity}\quad(\text{probe}=``+"\,,\ \text{pump}=``-")\,, \nonumber\\
    {\rm (B2):}&\quad\text{same helicity}\hspace*{6mm}\quad(\text{probe}=``+"\,,\ \text{pump}=``+")\,. \nonumber
\end{align}
Recall that in out setup in Fig.~\ref{fig:exp_setup} the propagation direction and the orientation of the electric field of the pump beam, and thus also its helicity, are reversed at the mirror acting as retro-reflector prior to the collision with the probe.
Finally, we assume (C) one beam to be linearly polarized and the other one to be circularly polarized. The specific cases considered here are
\begin{align}
    {\rm (C1):}&\quad\text{circ. polarized probe, lin. polarized pump}\quad(\text{probe}=``+"\,,\ \beta_{\rm pump}=\pi/2)\,, \nonumber\\
    {\rm (C2):}&\quad\text{lin. polarized probe, circ. polarized pump}\quad(\beta_{\rm probe}=\pi/2\,,\ \text{pump}=``-")\,. \nonumber
\end{align}

We emphasize that the polarizations of the colliding laser beams near focus encoded in $\vec{E}(x)$ certainly do differ from the polarizations of the collimated input beams. The reasons for this are both the focusing and the asymmetry induced by the use of the $90^\circ$ OAP in our setup. In addition, also the finite pulse duration of the colliding beams implemented by the above {\it ad hoc} description can impact the polarization of the beams. As we have explicitly ensured that $\omega\tau\gg1$, the latter effect should, however, be negligible.

In any case, we expect characteristic properties such as the dominant polarization component to be inherited by the focused beam.
To this end, it is, however, important to note that the $90^\circ$ reflection at the OAP genuinely maps the polarization vector of a linearly polarized input beam propagating in $-{\rm z}$ direction given by $\vec{e}_{\rm in}=\cos\beta\,\vec{e}_{\rm x}+\sin\beta\,\vec{e}_{\rm y}$ [cf. Eqs.~\eqref{eq:incidentfield} and \eqref{eq:inputlinpol}] onto $\vec{e}_{\rm out}=\cos\beta\,\vec{e}_{\rm z}-\sin\beta\,\vec{e}_{\rm y}$ for the output ray propagating along $+{\rm x}$ towards the focal point, i.e., the beam axis of the focused beam.
The reason for this is that upon hitting a mirror the electric field component tangential to the mirror surface is compensated by a field induced in the mirror such as to ensure the electric field to vanish identically on its surface.
Analogously, the polarization vector of a ``$\pm$'' circularly polarized input beam $\vec{e}_{\rm in}=(\vec{e}_{\rm x}\mp{\rm i}\,\vec{e}_{\rm y})/\sqrt{2}$ [cf. \Eqref{eq:inputcircpol}] is mapped onto $\vec{e}_{\rm out}=(\vec{e}_{\rm z}\pm{\rm i}\,\vec{e}_{\rm y})/\sqrt{2}$ by the $90^\circ$ OAP, which clearly implies that the reflection reverses the polarization from right to left hand circular polarization and vice versa.

At the same time, we emphasize that within the present approach we have direct access to the polarization vectors $\vec{e}_{\rm div}(\vec{k})$ of the laser photons of wavevector $\vec{k}$ forming the divergent laser beam propagating towards the retro-reflector in our setup in Fig.~\ref{fig:exp_setup}.
In line with this, we refer to signal photons that co-propagate with the divergent OAP reflected beam after focusing and fulfill the criterion
\begin{equation}
    \vec{e}_{p}(\vec{k}) \cdot \vec{e}_{\rm div}^{\,\,*} = 0 
    \label{eq:orthrel}
\end{equation}
as being perpendicularly polarized to this beam; we denote their polarization vectors by $\vec{e}_\perp(\vec{k})$.
For the details, see Appendix~\ref{sec:divlaserpolvec}.

\subsection{Prospective signals}

In Tab.~\ref{tab:resultshole} we provide our results for the numbers of signal photons reaching a detector registering all signal photons traversing the hole in the retro-reflector per shot, i.e., those signal photons scattered to polar angles $\vartheta\leq\theta_{\rm det}$.
To this end we integrate \Eqref{eq:photonnumberdens} over the full azimuthal angle $0\leq\varphi\leq2\pi$ and the polar angle interval $0\leq\vartheta\leq \arctan(r_{\rm det}/f_{\rm eff})\simeq 7.1^\circ$.
\begin{table}[h]
    \centering
    \begin{tabular}{lc||c|c|c|c|c|c|c|}
        & & (A1) & (A2) & (A3) & (B1) & (B2) & (C1) & (C2) \\
        \hline
        \hline
        \multirow{2}{*}{estimate\ } & $N_{\rm det}$ & 1.41 & 2.86 & 4.32 & 2.67 & 2.67 & 2.86 & 2.67 \\
        & $N_{\perp,{\rm det}}$ & 0 & 0.20 & 0 & 0 & 0 & 0.20 & 0 \\
        \hline
        \multirow{2}{*}{full calc.} & $N_{\rm det}$ & 1.48 & 2.88 & 4.29 & 3.11 & 2.32 & 2.88 & 2.72\\
        & $N_{\perp,{\rm det}}$ & $6.71 \times 10^{-4}$ & 0.19 & $7.78 \times 10^{-3}$ & $3.29 \times 10^{-11}$ & $2.72 \times 10^{-11}$ & 0.18 & $1.83 \times 10^{-2}$\\
        \hline
    \end{tabular}
    \caption{Numbers of signal photons scattered into the hole in the retro-reflector fulfilling $\vartheta\leq\theta_{\rm det}$. $N_{\rm det}$: polarization insensitive measurement; $N_{\perp,{\rm det}}$: perpendicularly polarized signal.}
    \label{tab:resultshole}
\end{table}
The estimates presented here are obtained from \Eqref{eq:Npapprox} which, as detailed in Sec.~\ref{sec:expsetup} above, is severely based on the paraxial approximation and completely neglects the asymmetry introduced by the $90^\circ$ OAP visible in Fig.~\ref{fig:probeanamoll}, as well as any effects associated with the finite Rayleigh length of the probe.
In turn, deviations from the results of a full numerical calculation in the parameter regime where $\theta_{\rm out}\simeq r_{\rm beam}/f_{\rm eff}\simeq0.47$ are to be expected.
Still, Tab.~\ref{tab:resultshole} shows that the analytical estimates~\eqref{eq:Npapprox} with coefficients~\eqref{eq:c_lin}-\eqref{eq:c_circlin} encoding the polarization dependence of the signal nevertheless predict the full numerical results as well as their general trends reasonably well.

This is in particular true for the numbers $N_{\rm det}$ of signal photons to be registered by the detector in a polarization insensitive measurement. In general, the analytical estimates deviate from these only by factors of $\approx0.9\ldots1.2$.
For the linearly polarized cases (A) the estimates and the results of the full calculation fulfill $N_{\rm det}|_\text{(A1)}<N_{\rm det}|_\text{(A2)}<N_{\rm det}|_\text{(A3)}$. Moreover, the ratios of $N_{\rm det}|_\text{(A2)}/N_{\rm det}|_\text{(A1)}\approx2$ and $N_{\rm det}|_\text{(A3)}/N_{\rm det}|_\text{(A2)}\approx1.5$ inferred from both approaches agree well with each other.
Also the degeneracy of the results for (A2) and (C1) that is to be expected from the estimate is reproduced by the full calculation.
At the same time, the outcomes for (B1), (B2) and (C2) agree for the estimate but differ for the full calculation.
As to be expected from the paraxial limit (cf. the second to last paragraph of Sec.~\ref{subsec:theory}), for the cases considered here the maximum signal of $N_{\rm det}\simeq4.29$ signal photons per shot is obtained for orthogonally polarized probe and pump beams in (A3).

\begin{figure}[h]
\begin{minipage}[t]{0.5\textwidth}
  \centering
  \includegraphics[width=\linewidth]{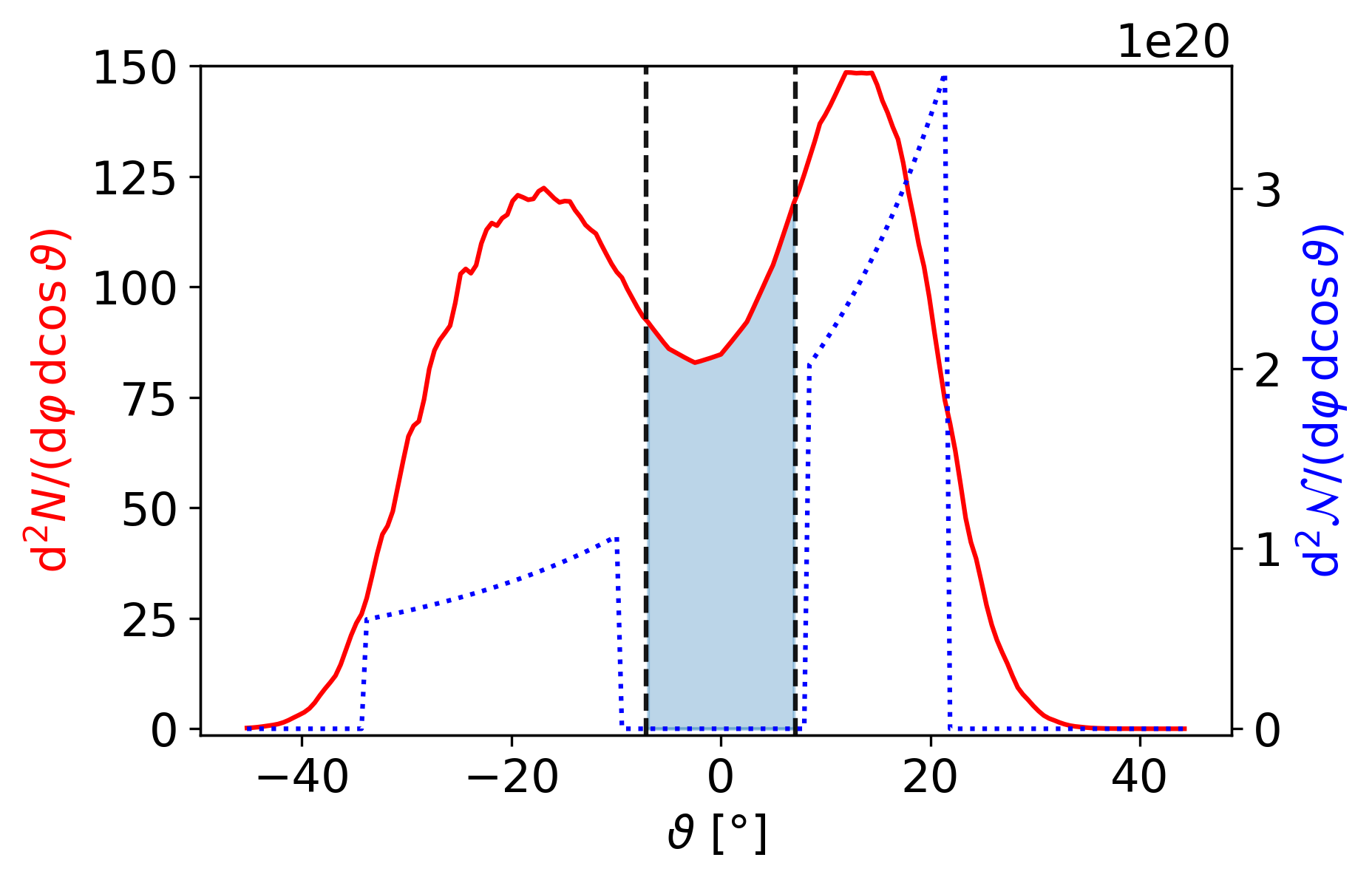}
\end{minipage}%
\begin{minipage}[t]{.51\textwidth}
  \centering
  \includegraphics[width=\linewidth]{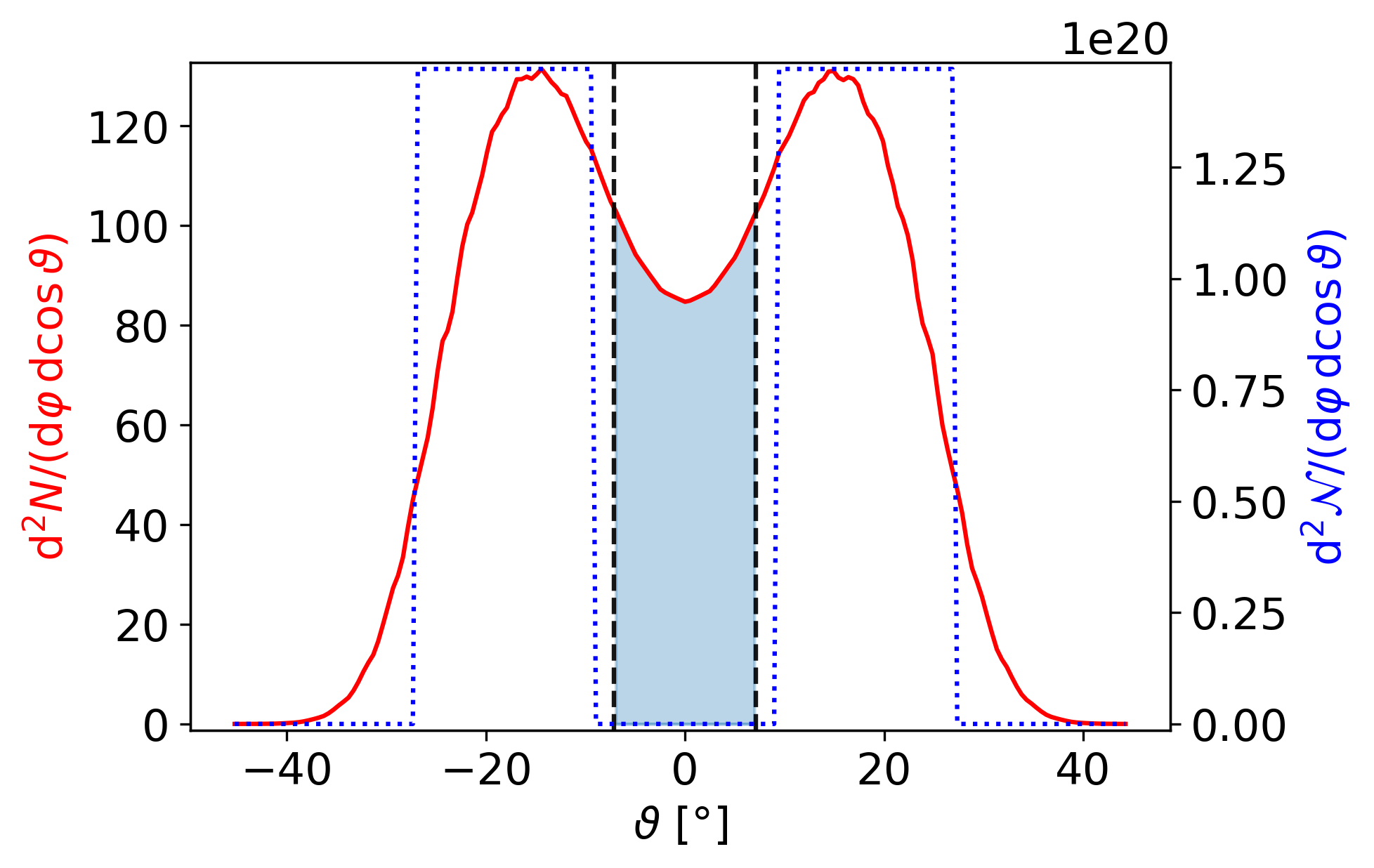}
\end{minipage}
\caption{Differential number~\eqref{eq:photonnumberdens} of signal photons $N$ attainable in a polarization insensitive measurement (solid red line) and laser photons $\cal N$ (dotted blue line) for case (A3). Dependence on $\vartheta$ for $\varphi=0$ (left) and $\varphi=\pi/2$ (right). The blue shaded areas delimited by dashed vertical lines mark the fraction of the signal reaching the detector.}
\label{fig:(A3)}
\end{figure}

In Fig.~\ref{fig:(A3)} we depict the angular emission characteristics of the polarization insensitive signal for this particular case;  note that ${\rm d}^2N_p(\varphi,-\vartheta)={\rm d}^2N_p(\varphi+\pi,\vartheta)$ and similarly for ${\rm d}^2{\cal N}_p$.
The depicted angular resolved signal photon distributions exhibit a pronounced dent in the field free shadow imprinted in the probe beam. We note that this behavior can be attributed to the specific collision geometry considered here where both the probe and the pump beam are focused to the same waist size. In principle, a peaked signal photon distribution in the shadow can be ensured but requires to tune the probe and pump waists appropriately \cite{Karbstein:2020gzg}.

For completeness, we also note that the full calculation allows us to infer that the total numbers $N$ of induced signal photons scattered into the right half-space, i.e., to angles $0\leq\varphi\leq2\pi$ and $0\leq\vartheta\leq\pi/2$, are about a factor of $\approx17.2\ldots17.8$ larger than those for $N_{\rm det}$ in Tab.~\ref{tab:resultshole}.
However, these numbers are typically only of academic interest as most of the signal photons are vastly dominated by the background of the driving laser photons and thus essentially inaccessible in experiment.

\begin{figure}
\begin{minipage}[t]{0.45\textwidth}
  \centering
  \includegraphics[width=1\linewidth]{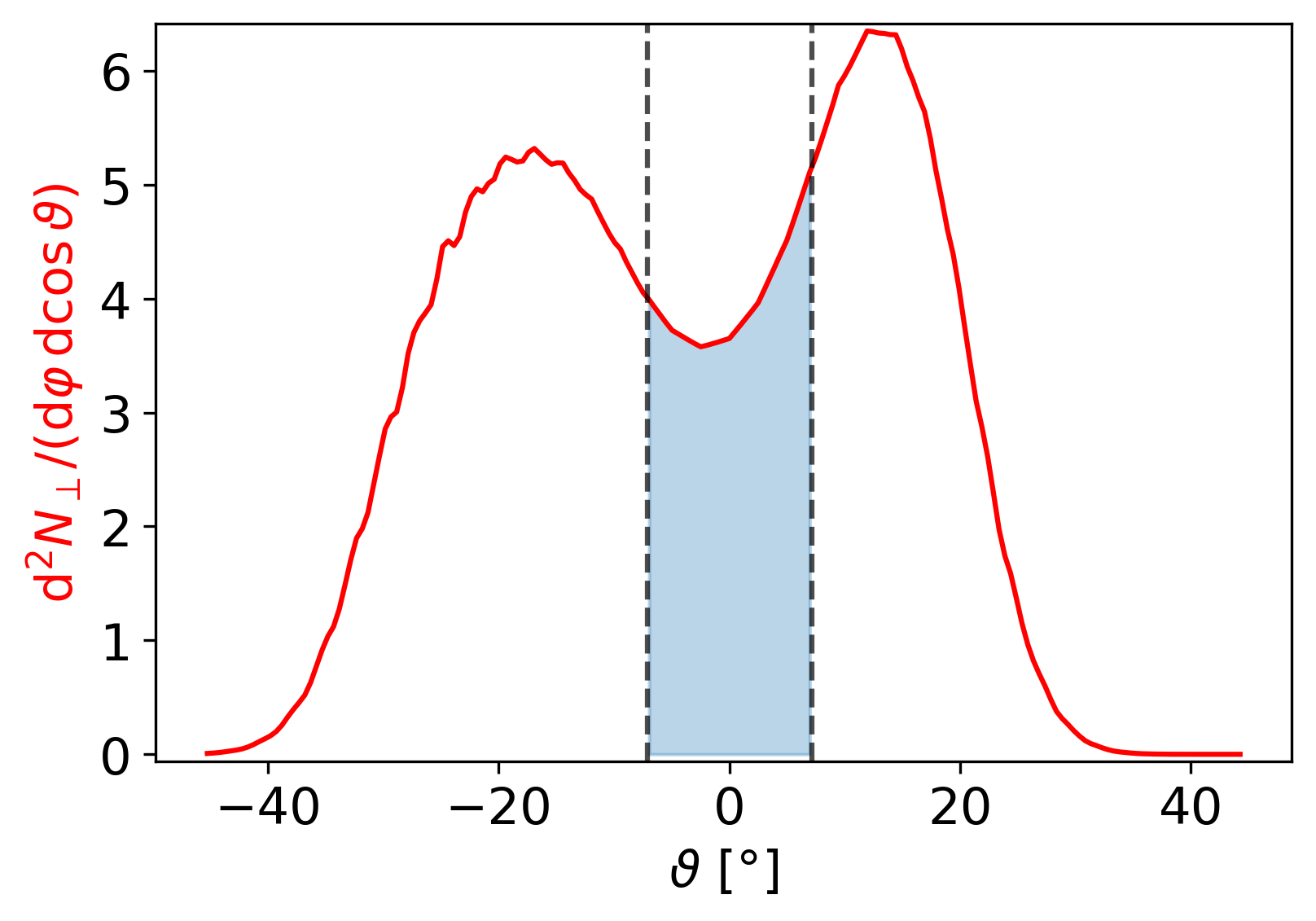}
\end{minipage}
\begin{minipage}[t]{.45\textwidth}
  \centering
  \includegraphics[width=1\linewidth]{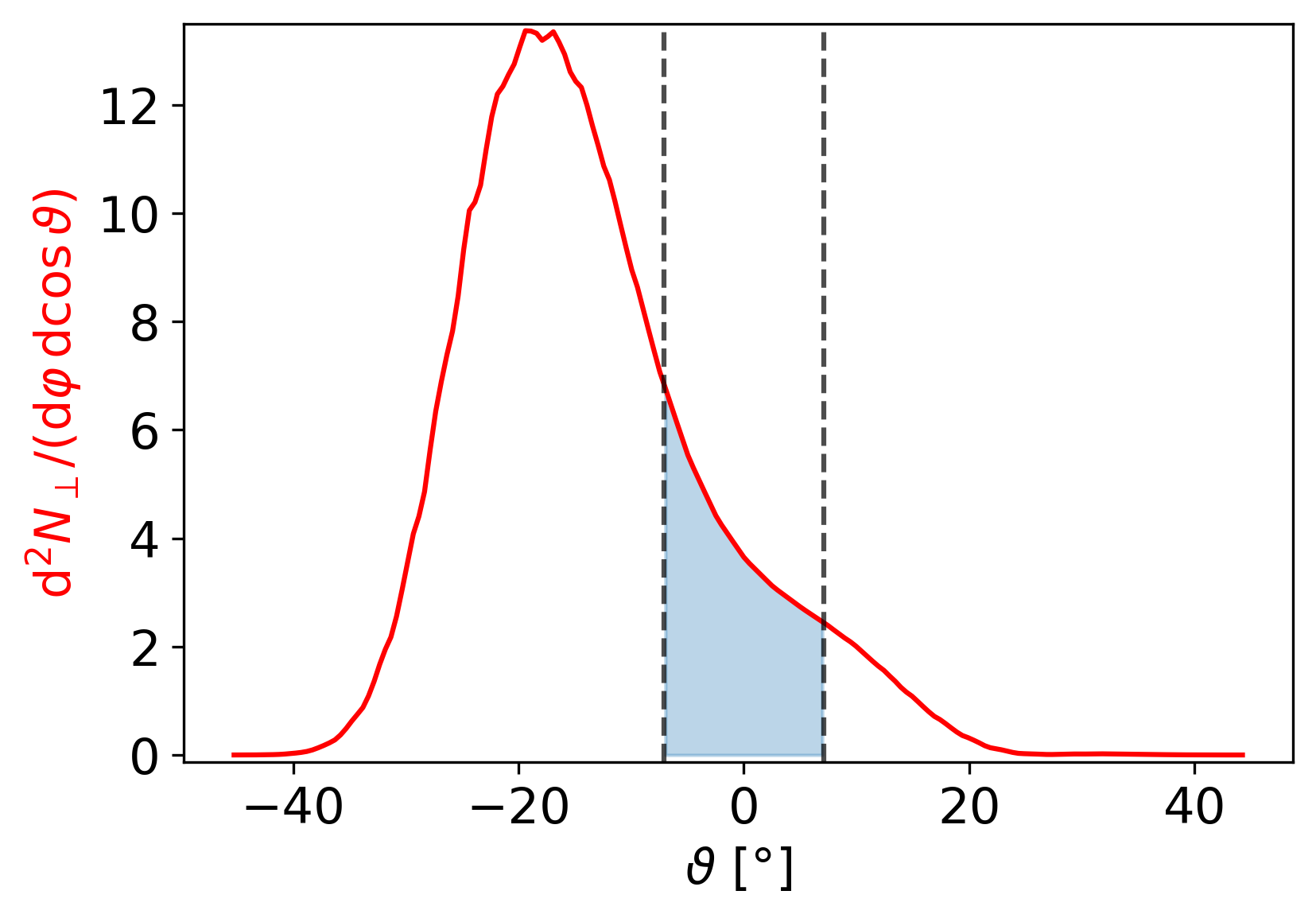}
\end{minipage}
\caption{Differential number~\eqref{eq:photonnumberdens} of $\perp$ polarized signal photons $N_\perp$ (solid red line) for case (A2). Note, that by definition there are no $\perp$-polarized laser photons. Dependence on $\vartheta$ for $\varphi=0$ (left) and $\varphi=\pi/2$ (right). The blue shaded areas delimited by dashed vertical lines mark the fraction of the signal reaching the detector.}
\label{fig:(A2)perp}
\end{figure}

For all cases apart from (A2) and (C1) the polarization-flipped signals per shot $N_{\perp,{\rm det}}$ are suppressed by at least two orders of magnitude relatively to the polarization unresolved signals $N_{\rm det}$.
This is in good accordance with the analytic estimates predicting $N_{\perp,{\rm det}}$ to vanish identically in these cases.
On the other hand, the analytical estimates for (A2) and (C1) overestimate the results of the full calculation for our setup by a factor of just $\approx1.1$. This minor discrepancy is in line with the one found above for the polarization insensitive signals $N_{\rm det}$.
Because (A2) maximizes the polarization-flip signal in the collision of linearly polarized beams, this choice is typically envisioned for vacuum birefringence experiments; cf., e.g., Refs.~\cite{Heinzl:2006xc,DiPiazza:2006pr}.
In accordance with the corresponding analytical estimates, the full calculation predicts the signal photon numbers $N_{\perp,{\rm det}}$ to (approximately) agree for (A2) and (C1) and result in a maximum yield of $N_{\perp,{\rm det}}\simeq0.19$ polarization-flipped signal photons per shot. Also note that for these cases the total-to-flipped signal ratio $N_{\rm det}/N_{\perp,{\rm det}}$, predicted to be $\simeq14.3$ by the analytical estimate, is found to be $\approx15.5\ldots15.8$ for the full calculation, and thus is in reasonable agreement.
Figure~\ref{fig:(A2)perp} shows the angular emission characteristics of the $\perp$ polarized signal for (A2).

\section{Conclusions and Outlook}\label{sec:concls}

In this work we have put forward a concrete experimental schema allowing to detect nonlinear quantum vacuum signals in the head-on collision of two tightly focused high-intensity beams of the petawatt class with state-of-the-art technology.

The key components of our setup are a $90^\circ$ off-axis parabolic mirror and a retro-reflector arranged such as to allow two subsequent laser pulses generated by the same frontend to be focused by the same optics and to be collided in the common focal point of the OAP and the retro-reflector. In order to allow for a measurement of the small quantum vacuum signal in the presence of the large background of the laser photons constituting the driving beams our setup employs a dark-field approach: to this end a central shadow is imprinted into the colliding laser beams and the signature of quantum vacuum nonlinearity to be detected in experiment amounts to signal photons scattered into the central shadow. The latter is imaged onto a detector through a hole in the retro-reflector. By appropriately preparing the polarization state of the incident beams prior to being fed into the focusing optics also polarization sensitive observables can be studied with our setup.

Resorting to a set of well-justified assumptions and theoretical idealizations, in the present work we have explicitly demonstrated that the setup envisioned by us should indeed provide a prospective new route towards a first measurement of nonlinear quantum vacuum signals in an all-optical experiment at present and forthcoming petawatt-class high-intensity laser laboratories, such as CALA in Garching, Germany.
As a critical next step, it needs to be shown that the scattering and diffraction background that is inevitable in any real-world experimental implementation of the setup can indeed be appropriately controlled and sufficiently suppressed in experiment.

\acknowledgments

This work has been funded by the Deutsche Forschungsgemeinschaft (DFG) under Grant Nos. 416607684, 416702141 and 416708866 within the Research Unit FOR2783/2.

\appendix

\section{Components of $\vec{E}(\omega,\vec{k})$ and far-field polarization vectors}\label{sec:E(omega,k)}\label{sec:divlaserpolvec}

Equations~\eqref{eq:E_OAP} and \eqref{eq:Fourierint} imply that the components of $\vec{E}(\omega,\vec{k})$ can be expressed as
\begin{align}
    E_{\rm x}(\omega,\vec{k}) &= \frac{\rm i}{\omega}(2\pi)^2\,\delta({\rm k}-\omega)\,\frac{f_{\rm eff}}{(\omega+k_{\rm z})^2}\bigl[E_{0,{\rm x}}(\omega^2+\omega k_{\rm z}-k_{\rm x}^2)-E_{0,{\rm y}}k_{\rm x}k_{\rm y}\bigr]\,, \nonumber\\
    E_{\rm y}(\omega,\vec{k}) &= \frac{\rm i}{\omega}(2\pi)^2\,\delta({\rm k}-\omega)\,\frac{f_{\rm eff}}{(\omega+k_{\rm z})^2}\bigl[E_{0,{\rm y}}(\omega^2+\omega k_{\rm z}-k_{\rm y}^2)-E_{0,{\rm x}}k_{\rm x}k_{\rm y}\bigr]\,, \nonumber\\
    E_{\rm z}(\omega,\vec{k}) &=- \frac{\rm i}{\omega}(2\pi)^2\,\delta({\rm k}-\omega)\,\frac{f_{\rm eff}}{\omega+k_{\rm z}}\bigl(E_{0,{\rm x}}k_{\rm x}+E_{0,{\rm y}}k_{\rm y}\bigr)\,,
    \label{eq:Eomegak}
\end{align}
where the arguments $\tilde{\rm x}$ and $\tilde{\rm y}$ of $E_{0,{\rm x}}$ and $E_{0,{\rm y}}$ are to be identified with \Eqref{eq:tildexy}.
One can easily convince oneself that the electric field components in \Eqref{eq:Eomegak} indeed fulfill $\vec{k}\cdot\vec{E}(\omega,\vec{k})=0$ in accordance with Maxwell's equations in vacuum.
The unit vector
\begin{equation}
 \vec{e}_{\rm div}(\vec{k})=\frac{1}{\bigl|\vec{E}(\omega,\vec{k})\bigr|}\,\vec{E}(\omega,\vec{k})\bigg|_{\omega=|\vec{k}|}
 \label{eq:eout}
\end{equation}
can hence be interpreted as the far-field polarization vector of the photons of wavevector $\vec{k}$ constituting the divergent laser beam after focusing. In our setup in Fig.~\ref{fig:exp_setup} these are the laser photons propagating towards the retro-reflector. The latter is sufficiently separated from the focal point F, such as to be effectively located in the far field; cf. also the corresponding discussion in the paragraph below \Eqref{eq:E0} in Sec.~\ref{sec:lasermodel}.
In line with that, the polarization vector of photons propagating towards the retro-reflector and being polarized perpendicular to the laser photons can be defined as
\begin{equation}
    \vec{e}_\perp(\vec{k}):=\vec{k}\times\vec{e}^{\ \!*}_{\rm div}(\vec{k})\,.
    \label{eq:eperp}
\end{equation}
Finally, some comments on a subtlety are in order here: clearly, \Eqref{eq:Eomegak} is non-zero only for directions $\vec{k}/|\vec{k}|$ for which the components $E_{0,{\rm x}}$ and $E_{0,{\rm y}}$ determined by the input field~\eqref{eq:incidentfield} [cf. also Eqs. \eqref{eq:d2Nlaser} and \eqref{eq:tildexy}] do not vanish.
This immediately implies that, even though the dependence on $E_{0,{\rm x}}$ and $E_{0,{\rm y}}$ drops out completely for our specific profile choices~\eqref{eq:inputlinpol}-\eqref{eq:inputcircpol}, \Eqref{eq:eout} is {\it a priori} only defined for these directions. 
However, we emphasize that for the present analysis it indeed amounts to a natural choice to extend the definitions~\eqref{eq:eout} and \eqref{eq:eperp} to arbitrary photon propagation directions.
This is completely in line with the original derivation of \cite{Bahk:2005} underlying our present considerations, where the transverse direction dependence of $E_{0,{\rm x}}$ and $E_{0,{\rm y}}$ effectively only acts as a regulator enforcing a finite transverse beam profile, but does not at all affect the polarization characteristics relative to an infinitely extended monochromatic plane-wave input field.

\section{Numerical Implementation}\label{sec:numimpl}

The numerical evaluation of \Eqref{eq:amplitude} requires the tuning of eight numerical parameters. These are the extents $(L_t,L_{\rm x}, L_{\rm y}, L_{\rm z})$ parameterizing the space-time volume of the simulation box and the corresponding numbers of grid points $n_{(4)}=(n_t,n_{(3)})=(n_t,n_{\rm x}, n_{\rm y}, n_{\rm z})$.
The simulation box needs to be large enough such as to capture the interaction region of the colliding laser pulses where the quantum vacuum signals are induced.
In our setup in Fig.~\ref{fig:exp_setup} this amounts to the region around the focal point F, the transverse and longitudinal extents of which are controlled by the beam waist $w_0$ and the Rayleigh length ${\rm z}_{\rm R}$, respectively. The temporal extent is set by the overlap of the colliding pulses and is controlled by the pulse duration $\tau$.

\begin{figure}[h]
  \centering
  \includegraphics[width=0.8\linewidth]{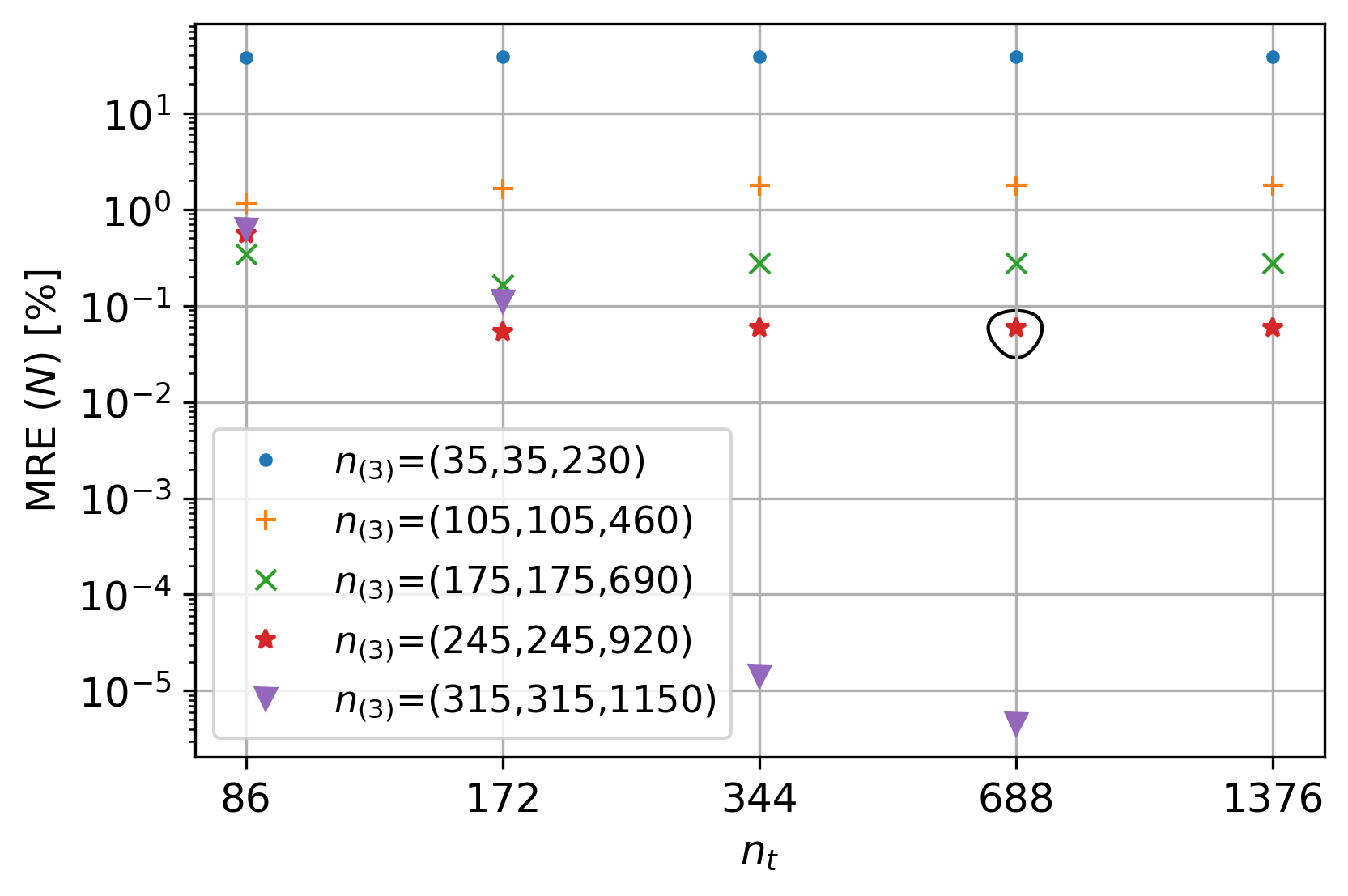}
  \caption{Log-log plot of the mean relative error (MRE) of the total number $N$ of signal photons scattered into the right half-space as a function of the number of temporal grid points $n_t$ for different spatial grid resolutions. The reference value for the MRE is $N \simeq 49.58$ obtained for the highest resolution of $n_{(4)} = (1376,315,315,1150)$. The data point enclosed by a circle (${\rm MRE} \simeq 0.06\,\%$) is determined with the default resolution of $n_{(4)}=(688,245,245,920)$ used in this work.}
  \label{fig:conv_test_Ntot}
\end{figure}

For the temporal extent we choose $L_t = 2\tau \simeq 102\,{\rm fs}$ centered at $t=0$ where the colliding pulses overlap best; recall that $\tau_{\rm FWHM}=30\,{\rm fs}$. 
Moreover, we choose $L_{\rm x} = 4c\tau \simeq 61.1\,\upmu{\rm m}$ and $L_{\rm y} = L_{\rm z} \simeq 19.1\,\upmu{\rm m}\approx24\lambda$ for the spatial extents in longitudinal and transverse directions, respectively.
We have explicitly checked that the field strengths at the boundaries reach at most $2\,\%$ of the peak field value for $t=0$.

A quantitative prediction of the quantum vacuum signals requires resolving all frequency components of the signal.
The four-field interaction in \Eqref{eq:HElagrangian} gives rise to signals at two different photon oscillation frequencies ${\rm k}\approx\{\omega,3\omega\}$.
In order to resolve the signal the maximum frequency ${\rm k}\approx3\omega$, we use $n_t = 688$ grid points in  temporal direction and $n_{\rm x} = 920$ grid points in longitudinal direction. This corresponds to a sampling with approximately $6$ and $4$ grid points per minimum wavelength $\lambda/3$, respectively. 
In the focal plane the annular flat top beam reflected off the $90^\circ$ OAP features side lobes that have to be resolved for an accurate determination of the signal.
As default value for our simulations we use $n_{\rm y} = n_{\rm z} = 245$ grid points in the transverse directions, which implies a sampling with approximately $3$ grid points per $\lambda/3$. For case (A2) in Sec.~\ref{subsec:laserpols}, the above extents of the simulation box and $n_{(4)}=(688,245,245,920)$ we obtain a total number of $N\simeq 49.58$ signal photons scattered into the right half-space.
With a finer resolution based on $n_{(4)}=(1376,315,315,1150)$ grid points the analogous result is found to be $N\simeq 49.61$. We use this value as reference value $N_{\rm ref}$ for the estimation of the mean relative error ${\rm MRE} = |N - N_{\rm ref}| / N_{\rm ref}$ of the signal photon numbers $N$ extracted from our simulations.
Therewith we attribute a ${\rm MRE}\simeq0.06\,\%$ to the results determined with the default resolution, indicating a reasonably small discretization error for the present purposes.
See Fig.~\ref{fig:conv_test_Ntot} for a plot of the MRE of $N$ as a function of $n_t$ for different spatial grid resolutions. 
Here, we observe a decrease of the MRE for increasing grid resolution.
Figure~\ref{fig:conv_test_Ntot} clearly indicates the convergence of the results for $N$ for the above finite space-time volume of the simulation box.

To discretize the signal photon energies and emission directions in spherical momentum coordinates, we use $n_{\rm k} = 718$ grid points in radial direction to span the energy range $0\leq{\rm k}\leq 14.5 \, {\rm eV}$, $n_\vartheta = 363$ grid points to discretize the azimuthal angle $0\leq\vartheta\leq\pi$ and $n_\varphi = 726$ grid points for the polar angle $0\leq\varphi\leq2\pi$.
This corresponds to a resolution of about $0.5^\circ$ for the angles and $0.02\,{\rm eV}$ for the energy.
To quantify the discretization error arising from the spherical momentum grid we also performed a reference calculation with $n_{\rm k}=1077$, $n_{\vartheta}=543$ and $n_{\varphi}=1086$ grid points. Therewith, we obtain a MRE of $\simeq 1.8\times 10^{-3}\,\%$ for $N$ and $\simeq 2.0\times10^{-3}\,\%$ for $N_{\perp}$ obtained by integration over $0\leq\varphi\leq2\pi$ and $0 \leq \vartheta \leq \pi/2$.

The accuracy of our numerical results does not only depend on the volume and the grid resolution of the simulation box just discussed, but of course also on the resolution of the discretization adopted for the numerical evaluation of \Eqref{eq:E_OAP}.
For the simulations presented in this work, we sample the rotationally symmetric beam of outer radius $r_{\rm beam}=14\,{\rm cm}$ in the input plane~\eqref{eq:incidentfield} by a rectangular grid of side lengths $l_{\rm x} = l_{\rm y} = 2r_{\rm beam}$ consisting of $n_{\rm x} = n_{\rm y} = 432$ points centered at the beam axis.
For comparison, we also performed a calculation at half resolution, i.e., for $l_{\rm x} = l_{\rm y} = 2r_{\rm beam}$ with $n_{\rm x} = n_{\rm y} = 216$ and extract the associated results for $N$ and $N_\perp$.
Calculating the MRE of these results with respect to the analogous values obtained at full resolution we find $\simeq 0.08\,\%$ for $N$ and $\simeq 0.20\,\%$ for $N_{\perp}$.
The quoted values for the MRE are of the same order as those inferred for the resolution of the simulation box used for the vacuum emission solver above. In turn, they also hint at a sufficiently small discretization error for the present work.

\end{document}